\documentclass[floats,floatfix,showpacs,amssymb,prd,twocolumn,superscriptaddress,nofootinbib]{revtex4-1}

\usepackage{anyfontsize}
\usepackage{amssymb}
\usepackage{amsmath}
\usepackage{amsfonts}
\usepackage{amsbsy}
\usepackage{newtxtext}
\usepackage{newtxmath}

\def\namedlabel#1#2{\begingroup
    #2%
    \def\@currentlabel{#2}%
    \phantomsection\label{#1}\endgroup
}

\usepackage{tensor}
\usepackage{mathrsfs}

\usepackage{bm}

\newcommand{\mpl}{{M_{\rm {pl}}}}

\newcommand{\dd}{\, {\rm d}}
\newcommand{\gsim}{\;\mbox{\raisebox{-0.5ex}{$\stackrel{>}{\scriptstyle{\sim}}$}
}\;}
\newcommand{\lsim}{\;\mbox{\raisebox{-0.5ex}{$\stackrel{<}{\scriptstyle{\sim}}$}
}\;}
\newcommand{\rdm}{{\rho_{\rm DM}}}

\newcommand{\tg}{\tilde{g}}

\newcommand{\nm}{{\mu\nu}}

\newcommand{\eff}{_{\rm eff}}

\newcommand{\ccc}{{_{\rm c}}}

\newcommand{\GN}{G_{\rm N}}

\newcommand{\T}{\Theta}

\usepackage[utf8]{inputenc}

\usepackage{graphicx}
\usepackage{epsfig}
\usepackage{epstopdf}

\usepackage[usenames,dvipsnames]{xcolor}

\usepackage{verbatim,mathtools,needspace,enumitem,etoolbox}

\definecolor{linkcolor}{rgb}{0.0,0.3,0.5}

\usepackage[unicode, colorlinks=true, linkcolor=linkcolor, citecolor=linkcolor, filecolor=linkcolor,urlcolor=linkcolor, pdfusetitle]{hyperref}

\usepackage{epstopdf}

\usepackage{bm}
\usepackage{dcolumn}

\usepackage{longtable}
 \usepackage[normalem]{ulem}

\definecolor{romared}{RGB}{142,0,28}
\hypersetup{colorlinks=true,
            citecolor=romared,
            linkcolor=romared,
            urlcolor=romared}
\def\aap{\rm{A\&A}}                

\def\mnras{\rm{MNRAS}}             
\begin{document}
\title{Screened Fifth Forces Mediated by Dark Matter--Baryon Interactions:\\ Theory and Astrophysical Probes}

\author{Jeremy Sakstein}
\email{sakstein@physics.upenn.edu}
\affiliation{Center for Particle Cosmology,
Department of Physics and Astronomy,
University of Pennsylvania,
209 S. 33rd St., Philadelphia, PA 19104, USA}

\author{Harry Desmond}
\email{harry.desmond@physics.ox.ac.uk}
\affiliation{Astrophysics, University of Oxford, Denys Wilkinson Building, Keble Road, Oxford OX1 3RH, UK}

\author{Bhuvnesh Jain}
\email{bjain@physics.upenn.edu}
\affiliation{Center for Particle Cosmology,
Department of Physics and Astronomy,
University of Pennsylvania,
209 S. 33rd St., Philadelphia, PA 19104, USA}

\raggedbottom

\begin{abstract}
  We derive the details of a new screening mechanism where the interactions of baryons and dark matter can be screened according to the local dark matter density. In this mechanism, the value of Newton's constant is dark matter density-dependent, allowing for the possibility that astrophysical phenomena are very different in galaxies less dense than the Milky Way. The parameterized post Newtonian parameter $\gamma$, which quantifies the difference between kinematical and lensing probes, also depends on dark matter density. We calculate the effects of varying $G$ on various stages of stellar evolution, focusing on  observables that impact cosmology: the Cepheid period--luminosity relation and the supernova Ia magnitude--redshift relation. Other potential tests of the model are also investigated including main-sequence, post-main sequence, and low mass dwarf stars. Finally, we discuss how extragalactic tests of $\gamma$ could provide complementary constraints.
\end{abstract}

\date{\today}

\maketitle

\section{Introduction}

Screening mechanisms play a paramount role in modern cosmology. Using non-linear effects, they act to suppress potential deviations from the standard $\Lambda$CDM cosmological model on small scales by hiding the new or \emph{fifth} forces that are ubiquitous in theories of dark energy that include new degrees of freedom (typically a light scalar) coupled to matter \cite{Clifton:2011jh,Joyce:2014kja,Koyama:2015vza,Sakstein:2015oqa,Burrage:2016bwy,Burrage:2017qrf,Sakstein:2018fwz,Ishak:2018his}. Typically, they are efficient in high density regions such as the Solar System and the Milky Way but do not operate at low densities in order to allow for novel cosmological phenomenologies. Models of dark energy that do not include screening mechanisms are typically fine-tuned or ruled out (see \cite{Khoury:2018vdv,Heckman:2019dsj} for some exceptions). This has made screening mechanisms the focus of much theoretical study \cite{Joyce:2014kja}, the inspiration for several laboratory tests \cite{Burrage:2016bwy,Burrage:2017qrf} specifically designed to search for them, and targets for upcoming cosmological surveys. 

The most-commonly studied mechanisms focus on screening interactions between a hypothetical new scalar and matter, although new vectors can also have screening mechanisms \cite{DeFelice:2016cri}, as can theories where the graviton is massive \cite{deRham:2016nuf}. Focusing on scalars\footnote{The screening mechanisms in vector and massive gravity theories can be phrased in terms of equivalent scalar screening mechanisms using either the St\"{u}ckelberg trick or by considering the decoupling limit.}, present screening mechanisms fall into one of two categories. The chameleon \cite{Khoury:2003aq,Khoury:2003rn}, symmetron \cite{Hinterbichler:2010es}, and dilaton \cite{Brax:2010gi} models screen by suppressing the new scalar charge, which quantifies an object's response to a fifth force field. Conversely, Vainshtein \cite{Vainshtein:1972sx,deRham:2016nuf} and K-mouflage \cite{Babichev:2009ee} screening suppresses the fifth force field compared with the Newtonian one.

Any new screening mechanism would have profound consequences for cosmology and gravitation because it would allow for the construction of novel non-trivial dark energy models and modified gravity theories. A new screening mechanism is apposite now more than ever, since the binary neutron star merger with optical counterpart has ruled out a plethora of dark energy models that utilize those mechanisms \cite{Sakstein:2017xjx,Creminelli:2017sry,Baker:2017hug,Ezquiaga:2017ekz}. Others, in particular chameleon, symmetron, and dilaton models, are already constrained to levels where they cannot explain dark energy \cite{Wang:2012kj}.

It is widely believed that no new screening mechanism exists, based on the following argument. Consider a scalar $\phi$ coupled to matter represented by an energy-momentum tensor $T_{\mu\nu}$ and perturb both about their background values so that $\phi=\bar\phi+\delta\phi$, $T_{\mu\nu}=\bar T_{\mu\nu}+\delta T_\nm$. The most general effective action for the perturbations is (ignoring disformal couplings, which do not give rise to new screening mechanisms \cite{Sakstein:2014isa,Ip:2015qsa}):
\begin{equation}
\label{eq:genscreen}
\delta S=\int\dd^4x \left[Z^{\mu\nu}(\bar\phi)\partial_\mu\delta\phi\partial_\nu\delta\phi-m_{\rm eff}(\bar\phi)\delta\phi^2+\beta(\bar\phi)\frac{\delta\phi}{\mpl}\delta T\right],
\end{equation}
where $\delta T=\eta^\nm\delta T_\nm$ is the trace of the perturbed energy-momentum tensor. This includes a non-canonical kinetic term $Z^{\mu\nu}$, an effective mass $m_{\rm eff}(\bar\phi)$, and a coupling to matter $\beta(\bar\phi)$, all of which are background-dependent. This action gives rise to a scalar-mediated fifth force
\begin{equation}\label{eq:f5gen}
    F_5\sim \frac{2\beta^2(\bar\phi)GM}{\sqrt{Z}r^2}e^{-m_{\rm eff}(\bar\phi)r},
\end{equation}
where $Z\sim\textrm{det}(Z^\nm)$. Upon inspection, one can discern three possible methods of suppressing this environmentally. One possibility is that the Compton wavelength $\lambda_{\rm C}\sim m_{\eff}(\bar\phi)^{-1}$ is smaller than a micron so that current laboratory tests are satisfied \cite{Adelberger:2003zx,Adelberger:2006dh,Adelberger:2009zz}. This is how the chameleon mechanism operates. Another is that the coupling to matter is small enough to evade fifth force searches in the Solar System \cite{Sakstein:2017pqi}. This is how the symmetron and dilaton mechanisms screen. The final is to have a large kinetic matrix $Z$, as is the case with the Vainshtein and K-mouflage mechanisms. The argument that there are no new screening mechanisms is that any new scalar interactions (either self or with matter) will ultimately be encapsulated in Eq.~\ref{eq:f5gen}.

Recently, the authors of Ref. \cite{Berezhiani:2016dne} have devised a dark energy model where the cosmic acceleration is driven not by new dynamical degrees of freedom but by baryon--dark matter interactions. In particular, the \emph{Jordan frame metric}, $\tg_\nm$, which baryons couple to, is a combination of the \emph{Einstein frame metric}, $g_\nm$, which couples to dark matter, and the dark matter four-velocity $u_\mu$:
\begin{equation}\label{eq:coupling}
    \tg_\nm=R^2(\rho_{\rm DM})\left(g_\nm+u_\mu  u_\nu\right)-Q^2(\rdm)\,u_\mu u_\nu.
\end{equation}
Here, $R$ and $Q$ are arbitrary functions that are assumed to tend to unity (so that $\tg_\nm=g_\nm$) at early times, or, equivalently, when the dark matter density is high, but become important around the present epoch and drive the cosmic acceleration without any need for dark energy. The authors speculated that any deviations from general relativity (GR) in high dark matter density environments would be highly suppressed as a necessary corollary of $R$ and $Q$ tending to unity at early times/high densities. This circumvents the argument that there cannot be new screening mechanisms because there are no new degrees of freedom in this model, {only the standard model, GR, and dark matter}.

{The reader may recognize the coupling of Eq.~\eqref{eq:coupling} as a disformal transformation of the metric. It is well-known that disformal transformations of this type can give rise to higher-derivative and non-linear derivative interactions --- Horndeski and its extensions --- when one inverts the transformation to work in the Jordan frame \cite{Bettoni:2013diz,Zumalacarregui:2013pma,Gleyzes:2014dya,Bettoni:2015wta} so this reader may wonder if our mechanism is one of the known mechanisms masquerading in the Einstein frame. This is not the case. One can see this by noting that Horndeski (and its extensions) only admit an Einstein frame if the effects of the non-linearities can be removed by a disformal transformation \cite{Bettoni:2013diz}. If one were to construct the Jordan frame action then all of the non-linear terms would cancel in the equations of motion. Another way to see that our mechanism is unique is to note that our derivation is linear in the new field --- {by which we mean that the screening does not rely on the solution to a non-linear equation for the dark matter field sourced by an object} --- whereas other screening mechanisms rely crucially on non-linearities to operate.}

In this work, we study this theory in detail and verify that a novel screening mechanism indeed exists. In particular, we will show that, as a consequence of the interactions,
\begin{enumerate}
    \item \label{it:G} Newton's constant becomes dark matter-density dependent, i.e. $G=G(\rdm)$, and
    \item \label{it:g} The parameterized post-Newtonian (PPN) parameter $\gamma$ becomes dark matter-density dependent. 
\end{enumerate}
Regardless of dark energy --- one can always consider theories with screened fifth forces that are not cosmologically important, as is the case with a large portion of the chameleon and symmetron parameter space as well as the quartic and quintic galileons and generalizations thereof --- a basic requirement for any theory producing \ref{it:G} and \ref{it:g} is that the Solar System is screened, otherwise stringent tests of gravity already performed would be violated \cite{Will:2004nx,Sakstein:2017pqi}. It is likely that a vast portion of the Milky Way (MW) is also screened since several recent tests of gravity aimed at constraining fifth forces apply here \cite{Burrage:2017qrf,Sakstein:2018fwz}.
This implies that tests of the theory should focus on extragalactic observables, including galaxies in halos less dense than that of the MW, and voids (including the local void). To this end, in Secs.~\ref{sec:Ceph}--\ref{sec:WD} we will study the consequences of \ref{it:G} for astrophysical objects that can be observed in other galaxies, with the ultimate aim of identifying novel astrophysical probes of this theory. It is possible that some objects could be unscreened in our own galaxy in regions where the dark matter density is smaller than in the solar neighborhood. For this reason, we also discuss potential probes within the MW in Sec. \ref{sec:dwarf_stars}. Astrophysical tests have been incredibly successful for constraining other screening mechanisms \cite{Davis:2011qf,Jain:2012tn,Vikram:2014uza,Koyama:2015oma, Sakstein:2015zoa,Sakstein:2015aac, Sakstein:2016lyj, Babichev:2016jom, Sakstein:2016oel, Sakstein:2016ggl, Sakstein:2017bws, Sakstein:2017pqi,Adhikari:2018izo,Desmond:2018sdy,Desmond:2018euk,Desmond:2018kdn}. In Sec.~\ref{sec:ML} we discuss how \ref{it:g} could be used to further constrain the theory using extragalactic tests comparing lensing and kinematical tracers of the gravitational potential.

The layout of the paper is as follows. In Sec.~\ref{sec:DM} we review the model of Ref. \cite{Berezhiani:2016dne} and discuss its salient features. In Sec.~\ref{sec:SM} we derive the new screening mechanism, in particular \ref{it:G} and \ref{it:g}, and discuss possible tests. The main astrophysical test we will consider is devised in Sec.~\ref{sec:Ceph}. Here, we study the implications of \ref{it:G} for post-main-sequence stars with masses in the range $5M_\odot$--$13 M_\odot$. Stars with these masses may undergo Cepheid pulsations, and we derive the change in the period--luminosity relation (PLR) which is used to measure the distance to extragalactic hosts and calibrate the distance ladder. In Sec.~\ref{sec:MS} we analytically and numerically investigate the properties of unscreened main-sequence stars. As expected, they are more luminous than their screened counterparts due to the increased rate of nuclear burning needed to maintain hydrostatic equilibrium when their self-gravity is stronger. In Sec.~\ref{sec:TRGB} we study the tip of the red giant branch (TRGB) which is used as a distance indicator. We find that the TRGB distance to unscreened galaxies is over-estimated, and thus comparing TRGB and Cepheid distances (which under-estimate the distance) is a promising method of constraining our screening mechanism. In Sec.~\ref{sec:WD} we discuss white dwarfs and type Ia supernovae (SNe) in unscreened galaxies. The Chandrasekhar mass is smaller when the value of $G$ is increased, while the rescaled supernovae peak luminosity is larger. In Sec.~\ref{sec:dwarf_stars} we study low mass dwarf stars which could potentially be unscreened in the outskirts of the MW where the dark matter density is lower than it is locally. In particular, we derive the effects of increasing $G$ on the minimum mass for hydrogen burning (separating red and brown dwarfs) and the radius plateau in the brown dwarf Hertzsprung--Russell diagram. The minimum hydrogen burning mass decreases and the brown dwarf radius is smaller, as one would expect for more compact objects. In Sec.~\ref{sec:ML} we discuss possible tests of \ref{it:g} in the form of mass vs. light comparisons. Both strong and weak lensing of extragalactic sources can be used to probe $\gamma$ by comparing with suitable dynamical tracers such as the X-ray surface brightness or stellar velocity dispersion. Finally, we summarize our findings, discuss avenues for future work, and conclude in Sec.~\ref{sec:concs}. At times, we will make reference to polytropic models and the Eddington standard model: a brief review of these is provided in Appendix \ref{sec:poly}.

Although our tests are framed in terms of the baryon--dark matter interaction model, they are all designed simply to look for signatures of varying Newton's constant outside the solar system. Thus, they apply equally to any theory of gravity in which $G$ varies spatially.

\textbf{Conventions:} We use the mostly plus convention for the Minkowski metric, $\eta_\nm=\textrm{diag}(-1,\,1,\,1,\,1)$. We use $\GN$ to refer to the value of Newton's constant measured locally (in the Solar System) and $G(\rdm)$ to refer to its value at general $\rdm$. The Planck mass, which appears in the Einstein-Hilbert action, is $\mpl^2=(8\pi G)^{-1}$, i.e. $G$ without any stated density dependence is a constant. In all cases $\log(x)\equiv\log_{10}(x)$ unless otherwise stated.

\section{Dark Matter--Baryon Interactions}
\label{sec:DM}


Here we briefly review the model proposed by Ref. \cite{Berezhiani:2016dne}. In this model, baryon--dark matter interactions give rise to an effective space-time for the baryons that depends on the dark matter density. The dark matter is modelled as an irrotational fluid. The effective field theory for this is simply a $P(X)$ theory\footnote{In general, one needs three scalars $\Phi^I$ to describe a fluid but in the absence of vorticity there is only one degree of freedom and so the theory is dual to a $P(X)$ theory \cite{Dubovsky:2005xd}.}
for a scalar $\Theta$ with $X=-g^{\mu\nu}\partial_\mu\Theta\partial_\nu\Theta$. (Note that we take $\Theta$ to have dimensions of length and $P(X)$ to have dimensions of $[\textrm{mass}]^4$.) The action for this sector (including gravity) is
\begin{equation}\label{eq:action}
    S=\int\dd^4 x\sqrt{-g}\left[\frac{R(g)}{16\pi G}+P(X)\right],
\end{equation}
which gives the energy-momentum tensor for dark matter
\begin{equation}
    T^{\rm DM}_{\mu\nu}=2P_X\partial_\mu\Theta\partial_\nu\Theta+P g_{\mu\nu},
\end{equation}
where $P_X\equiv\dd P(X)/\dd X$. From this, we can identify the four-velocity, pressure, and density of the dark matter:
\begin{equation}
    u_\mu=-\frac{\partial_\mu\Theta}{\sqrt{X}},\quad P_{\rm DM} = P(X),\quad\textrm{and}\quad \rho_{\rm DM}=2P_XX-P(X).
\end{equation}
Since we are interested in pressureless non-relativistic dark matter we can take $P_{\rm DM}\ll\rho_{\rm DM}$, which implies that
\begin{equation}\label{eq:DMdens}
    \rho_{\rm DM}\approx 2P_XX.
\end{equation}

The baryon--dark matter interactions arise because the baryons move on an effective \emph{Jordan frame} metric $\tilde{g}_{\mu\nu}$ that is a combination of the \emph{Einstein frame} metric $g_{\mu\nu}$ and the dark matter variables. In particular
\begin{align}
\tilde{g}_{\mu\nu}&=R^2(X)(g_{\mu\nu}+u_\mu u_\nu)-Q^2(X)u_\mu u_\nu \nonumber\\&= R^2(X)g_\nm + S(X)\partial_\mu\Theta\partial_\nu\Theta,
\end{align}
where $R(X)$ and $Q(X)$ are free functions and
\begin{equation}
    S(X)\equiv \frac{R^2(X)-Q^2(X)}{X}.
\end{equation}
The baryon--dark matter interaction is incorporated in the model by augmenting the action \eqref{eq:action} with a term 
\begin{equation}
    S_\textrm{b}[\tilde{g},\{\psi_{\textrm{b},i}\}]=\int\dd^4x\sqrt{-\tilde{g}}\mathcal{L}_b(\tilde{g},\{\psi_{\textrm{b},i}\}),
\end{equation}
where $\psi_{\textrm{b},i}$ represent the various baryon fields. Cosmic acceleration is achieved as follows. Considering a flat Friedmann-Lema\^{i}tre-Robertson-Walker Universe given by
\begin{align}
    \dd s^2=-\dd t^2+ a^2(t)\dd\vec{x}^2
\end{align}
in the Einstein frame, the field $\Theta$ is time-dependent so that, using Eq. \eqref{eq:DMdens}, any factor of $X$ or $\dot{\Theta}$ can always be replaced by a function of $\rdm(t)$. This metric is decelerating since at late times $a(t)\sim t^{2/3}$ (recall there is no dark energy in this model). The Jordan frame metric is then
\begin{equation}
    \dd\tilde{s}^2=-Q^2(\rdm)\dd t^2+R^2(\rdm)a^2(t)\dd\vec{x}^2.
\end{equation}
The functions $R$ and $Q$ are chosen such that, at late times i.e. $\rdm\sim H_0^2/\GN$, the scale factor in this frame, $\tilde{a}(\tilde{t})=R(\rdm(t(\tilde{t}))a(t(\tilde{t}))$, is accelerating. (The time-coordinate has been changed from $t$ to $\tilde t$ by defining $\dd \tilde{t}=Q(\rdm)\dd t$.) Thus, cosmic acceleration is achieved without the need for any dark energy degree of freedom.
At early times, the standard cosmological history is recovered by assuming that both $R$ and $Q$ tend to unity so that $\tg_\nm\rightarrow g_\nm$. Since these functions describe the interaction of dark matter with baryons, this must be a density dependent effect that suppressed deviations from GR when $\rdm\gsim H_0^2/\GN$. This observation led the authors of \cite{Berezhiani:2016dne} to speculate that a similar mechanism would screen the interactions on smaller scales, inside collapsed objects. In the next section we show that this is indeed the case.

\section{The Screening Mechanism}
\label{sec:SM}

In this section we derive \ref{it:G} and \ref{it:g}, namely that Newton's constant and the PPN parameter $\gamma$ are density-dependent in this theory. (The reader interested only in astrophysical tests can safely skip to the subsequent sections.) We accomplish this by calculating the weak-field limit for an isolated star immersed in a background dark matter density. We assume that the dark matter density does not vary spatially over length scales shorter than the radius of the star. This means that we can treat $X$ as constant and perturb $\Theta=\bar{\Theta}-\theta$ so that $X=\dot{\bar{\T}}^2-2\dot{\bar{\T}}\dot\theta+\mathcal{O}(\theta^2)$. The four-velocity of the dark matter is then (ignoring post-Newtonian corrections)
\begin{equation}\label{eq:FV}
    u^\mu=\left(1+\mathcal{O}(\dot{\theta}^2),\,\frac{\partial^i\theta}{\dot{\bar\T}}\right),
\end{equation}
from which we can identify the dark matter velocity $v^i_{\rm DM}=\partial^i\theta/\dot{\bar{\T}}$. Since the dark matter is non-relativistic this is $\ll1$. We assume further that the star itself does not source an appreciable amount of dark matter (i.e. dark matter does not cluster inside the star) so that we simply need to solve the Einstein equations for a baryon source.
These are
\begin{equation}\label{eq:EE}
    G_\nm = 8\pi G T_{{\rm b},\,\mu\nu},
\end{equation}
where $T_{\textrm{b}}^{\mu\nu}=2/\sqrt{-g}\delta S_{\rm b}/\delta g_\nm$. Now since the baryons couple to dark matter this is not covariantly conserved ($\nabla_\mu T^{\mu\nu}_{\textrm{b}}\ne0$). It is the Jordan frame energy-momentum tensor $\tilde{T}_{\textrm{b}}^{\mu\nu}=2/\sqrt{-\tilde{g}}\delta S_{\rm b}/\delta \tilde{g}_\nm$ ($\sqrt{-\tilde{g}}=QR^3\sqrt{-g}$) that is covariantly conserved (with respect to the Jordan frame metric, $\tilde{\nabla}_\mu \tilde{T}^{\mu\nu}_{\textrm{b}}=0$). This means that quantities such as the baryon pressure and density should be defined in the Jordan frame using $\tilde{T}_{\textrm{b}}^{\mu\nu}$. They can then be used in Eq. \eqref{eq:EE} by converting to the Einstein frame energy-momentum tensor as follows \cite{Berezhiani:2016dne}:
\begin{equation}\label{eq:Ttilde}
    {T}_{\textrm{b}\,\mu\nu}=QR^3\tilde{T}_{\textrm{b}}^{\kappa\lambda}\left[R^2g_{\mu\kappa}g_{\mu\lambda}+(2RR_X g_{\kappa\lambda}+S_X\partial_\kappa\T\partial_\lambda\T)\partial_\mu\T\partial_\nu\T\right].
\end{equation}
In order to calculate this for our non-relativistic setup we need to specify a coordinate system. By virtue of our choice of four-velocity in Eq. \eqref{eq:FV} we are working in Minkowski space in Cartesian coordinates, i.e.
\begin{equation}
    \dd s^2=-\dd t^2 + \dd x^2
\end{equation}
at the background level. As we are only interested in calculating the weak-field limit to Newtonian order ($\mathcal{O}(v^2/c^2)$), it is sufficient to calculate the Jordan frame metric to zeroth-order to compute the energy-momentum tensor. At this order, one has
\begin{equation}
    \dd\tilde{s}^2=-Q^2\dd t^2 + R^2\dd x^2.
\end{equation}
This implies that
\begin{equation}
    \tilde{T}_{\textrm{b}}^{\mu\nu}=\textrm{diag}\left(\frac{\tilde{\rho}_{\textrm{b}}}{Q^2},\,0,\,0,\,0\right),
\end{equation}
where $\tilde{\rho}_{\textrm{b}}$ represents the Jordan frame density and we have assumed that the baryons are pressureless (i.e. the non-relativistic limit).

Now we are in a position to calculate the source for Eq. \eqref{eq:EE} using Eq. \eqref{eq:Ttilde}:
\begin{align}
   {T}_{\textrm{b}\,00}&=\frac{R^3}{Q}\left[R^2+\dot{\bar{\T}}^2\left(S_X\dot{\bar{\T}}^2-2R R_X\right)\right]\tilde{\rho}_{\textrm{b}},\\
    {T}_{\textrm{b}\,0i}&=-\frac{R^3}{Q}\left[S_X\dot{\bar{\T}}^2-2R R_X\right]\tilde{\rho}_{\textrm{b}}\dot{\bar{\T}}\partial_i\theta,\\
    {T}_{\textrm{b}\,ij}&=\frac{R^3}{Q}\left[S_X\dot{\bar{\T}}^2-2R R_X\right]\tilde{\rho}_{\textrm{b}}\partial_i\theta\partial_j\theta.
\end{align}
Since $\partial_i\theta\sim v_i$, the $0i$- and $ij$-components are at a higher post-Newtonian order than the $00$-component and we can safely neglect them\footnote{Note that there are two velocities here---$v_i$, the dark matter velocity, and the velocity of the baryon fluid---so really this is a post-Newtonian expansion in both velocities. The latter velocity is post-Newtonian by definition, so we are implicitly assuming that the velocity of the dark matter is of the same order or less than the baryon fluid velocity.}. The important point is that we can invert Eq. \eqref{eq:DMdens} to write $\bar{X}=\dot{\bar{\T}}^2$ and $\dot{\bar{{\T}}}$ as a function of $\rho_{\rm DM}$, which means that the source for the Einstein equations is a function of the local dark matter density. In particular, we can expand Einstein's equations assuming the Newtonian limit for the metric 
\begin{equation}
    g_\nm=(-1+2\Phi)\dd t^2 + (1+2\Psi)\delta_{ij} \dd x^i\dd x^j,
\end{equation}
to find
\begin{equation}\label{eq:PhiEF}
    \nabla^2\Phi=\nabla^2\Psi=-4\pi Gf(\rho_{\rm DM})\tilde{\rho}_{\textrm{b}}
\end{equation}
with
\begin{equation}
    f(\rho_{\rm DM})=\frac{R^3}{Q}\left[R^2+\dot{\bar{\T}}^2\left(S_X\dot{\bar{\T}}^2-2R R_X\right)\right].
\end{equation}
The solution of equation \eqref{eq:PhiEF} is 
\begin{equation}\label{eq:Phisol}
    \Phi=\Psi=f(\rho_{\rm DM})\frac{G}{r}\int \dd^3\vec{x} \tilde{\rho}_{\textrm{b}}.
\end{equation}
Typically, one would identify the integral with the star's mass $M$ (or, in the case of the interior, the mass enclosed within a sphere of radius $r$), but in our case there is a subtlety. Because $\tilde{\rho}_{\textrm{b}}$ is defined in the Jordan frame we actually have to use the volume element defined using the Jordan frame metric. For this reason, one has
\begin{equation}
    M=\int\dd^3\vec{x}R^3\tilde{\rho}_{\textrm{b}},
\end{equation}
and therefore Eq. \eqref{eq:Phisol} is
\begin{equation}\label{eq:Phisol2}
    \Phi=\Psi=\frac{f(\rho_{\rm DM})}{R^3}\frac{GM}{r}.
\end{equation}
Using this, we can calculate the Jordan frame metric\footnote{Note that there is a contribution to the $0i$-component of the form $\tilde{g}_{0i}=S\dot{\bar\T}^2v_{\textrm{DM}\,i}$. This can be moved to the $00$-component by performing a linearized gauge transformation $x'^\mu=x^\mu+\xi^\mu$ with $\xi_i=0$ and $\xi_0=-S\dot{\bar\T}^2v_{\textrm{DM}\,i}x^i$ \cite{Sakstein:2014isa,Ip:2015qsa}. This adds a term $-2S\dot{\bar\T}^2v_{\textrm{DM}\,i}v_{\rm b}^i$ to the $00$-component, which we have neglected. It would be interesting to investigate the effects of this (non-PPN) term for the dynamics of test bodies in this space-time. We leave this for future work.}
\begin{align}
   \dd\tilde{s}^2&=\tilde{g}_\nm \dd x^\mu \dd x^\nu\nonumber\\&=Q^2\left(-1+2\frac{f(\rho_{\rm DM})}{RQ^2}\frac{GM}{r}\right)\dd t^2\nonumber\\&+R^2\left(1+2\frac{f(\rho_{\rm DM})}{R^3}\frac{GM}{r}\right)\delta_{ij}\dd x^i\dd x^j.
\end{align}
The simplest way to extract the observable quantities is to change gauge to quasi-Cartesian coordinates. This can be accomplished by defining
\begin{equation}
   \tilde{t}=Qt\quad\textrm{and}\quad \tilde{x}^i=R x^i.
\end{equation}
One then finds
\begin{align}
   \dd\tilde{s}^2&=\left(-1+2\frac{G_{\rm N}(\rho_{\rm DM})M}{\tilde{r}}\right)\dd \tilde{t}^2\nonumber\\&+\left(1+2\gamma(\rho_{\rm DM})\frac{G_{\rm N}(\rho_{\rm DM})M}{\tilde{r}}\right)\delta_{ij}\dd\tilde{x}^i\dd\tilde{x}^j,
\end{align}
where the gravitational constant
\begin{equation}\label{eq:GNrho}
     G(\rho_{\rm DM})=\frac{f(\rho_{\rm DM})}{Q^2}G.
\end{equation}
We remind the reader that this is solely a function of $\rho_{\rm DM}$ because we can always express $\dot{\bar{\T}}$ in terms of this. We can also identify the PPN parameter
\begin{equation}\label{eq:gammarho}
    \gamma(\rho_{\rm DM})=\frac{Q^2}{R^2},
\end{equation}
which controls light bending and the Shapiro time-delay effect. It is also a function of $\rho_{\rm DM}$. GR predicts that this is unity and in the Solar System deviations are constrained to be smaller than $2\times10^{-5}$ by the Cassini measurement of the Shapiro time-delay \cite{Bertotti:2003rm}. Since we do not have similar measurements in lower-density unscreened galaxies this bound does not apply in general.

A few comments are in order. 
First, note that when the Jordan frame metric is \emph{maximally conformal}, i.e. $S=0$ ($Q=R$), one has $G_{\rm N}=R^2G$ and $\gamma=1$ as one would expect\footnote{One would have $\gamma\ne1$ if $R$ were a function of $\T$ rather than $X$ because the baryon fluid would source the scalar. Since we have assumed this is not the case we expect $\gamma=1$. See Ref. \cite{Burrage:2017qrf} for more details.}. Second, in the \emph{maximally disformal case} $R=1$ the results are identical to those derived for pure disformal couplings by Ref. \cite{Ip:2015qsa}. Finally, one could derive the other PPN parameters by extending the calculation here. We expect that this will be significantly more complicated due to the necessity of including both the dark matter velocity and post-Newtonian baryon fluid variables. It is also likely that one will need to perform a complicated gauge transformation once the Jordan frame metric has been found (see \cite{Ip:2015qsa}).

Equations \eqref{eq:GNrho} and \eqref{eq:gammarho} are the main results of this section. Ref. \cite{Berezhiani:2016dne} did not give canonical forms for $R$ and $Q$ such that they can drive the cosmic acceleration. Since we wish to remain agnostic, we too will not provide functional forms but, rather, will simply treat Newton's constant and $\gamma$ as functions of $\rdm$. This allows for the possibility of model-independent tests of the theory that can be used to constrain dark energy (or other) models if and when they are devised. Our only demands will be that $G(\rho_{\rm loc})=\GN$ and that $\gamma(\rho_{\rm loc})$ satisfies the Cassini bound, where $\rho_{\rm loc}$ is the dark matter density at the location of the Solar System in the Milky Way's halo, $\sim10^{7}M_\odot\textrm{ kpc}^{-3}$ \cite{Read, Buch}. This ensures that the Solar System is screened so that any tests of this theory are necessarily extragalactic, and must be performed using galaxies with dark matter halos that are less dense than the MW. Depending on the form of $G(\rdm)$, it is also possible that the outskirts of the MW is unscreened.

Having identified the novel screening mechanism, the rest of this paper is devoted to finding astrophysical probes of it. We speculate on other potential tests in Sec.~\ref{sec:future}. Since we require the MW to be screened, we focus on objects that are abundantly observed in other galaxies. We will focus primarily on Cepheid variable stars, both because they can be modelled numerically, and because their use as distance indicators has previously been successful in constraining screening mechanisms that only exhibit effects in other galaxies. We will also discuss white dwarf stars and type Ia supernovae, as well as main-sequence stars, tip of the red giant branch, and low mass stars (red and brown dwarfs).


In this work, we will take the threshold for screening to be the ambient dark matter in the solar system, $\rho_{\rm loc}=10^7 M_\odot$ kpc$^{-3}$\footnote{This is the cosmological dark matter density at $z\approx40$, which may imply that this model predicts that deviations from $\Lambda$CDM emerge around this epoch {or later since it is possible that the threshold dark matter density is lower than in the MW.} .}. Objects in regions denser than this have $G(\rdm)=\GN$ whereas those in regions less dense than this have $G(\rdm)>\GN$\footnote{Note that we have implicitly assumed that the fundamental theory functions are such that the strength of gravity is enhanced in low density environments. One could instead have $G<\GN$, in which case the results of the tests studied in this work would be opposite. For example, main-sequences stars would be less luminous.}, with $\Delta G$ a constant independent of $\rdm$. Since $G(\rdm)$ does not vary over the objects we consider, we can then derive the change in these objects by setting 
\begin{equation}
    G(\rdm)=\left(1+\frac{\Delta G}{\GN}\right)\GN.
\end{equation}
It is important to bear in mind that in general the value of $\Delta G$ is likely to vary between objects inhabiting regions of different dark matter densities.

\vspace{4mm}

\noindent The reader uninterested in the technical details of the tests should consult Table \ref{tab:tests}, and may then skip to Sec.~\ref{sec:concs}.

\renewcommand{\arraystretch}{2.0}
\begin{table*}[ht]
    \centering
    \small\addtolength{\tabcolsep}{-5pt}
    \begin{tabular}{|c|c|c|}\hline
        Probe & Effect & Tests  \\\hline
        Cepheid stars & Modification of Period--Luminosity relation & Compare distance indicators\\
        Main sequence stars & Luminosity $\propto G(\rdm)^3$ (low mass) or $G(\rdm)$ (high mass) & Hertzsprung--Russell diagram, globular clusters\\
        Tip of the red giant branch & Decrease in tip luminosity & Compare distance indicators \\
        Red dwarf stars & Minimum hydrogen burning mass $\propto G(\rdm)^{1.398}$ & Hydrogen burning lines in atmosphere\\
        Brown dwarf stars & Decrease in radius $\propto G(\rdm)^{-1/2}$ at fixed mass & Measurement of radius plateau\\
        Lensing (strong and weak) & More lensing per unit mass ($\gamma>1$) & Compare dynamical and lensing masses\\\hline
        
    \end{tabular}
    \caption{Summary of astrophysical probes of the screening mechanism (Secs.~\ref{sec:Ceph}--\ref{sec:ML}).}
    \label{tab:tests}
\end{table*}


\section{Cepheid Variable Stars and the Period--Luminosity Relation}
\label{sec:Ceph}

 Cepheid variable stars have proven to be useful tools for constraining screening mechanisms \cite{Jain:2012tn,Sakstein:2015oqa,Sakstein:2016lyj,Sakstein:2018fwz}. 
 In this section we introduce Cepheids and the PLR, and calculate the change to both when Newton's constant is varied. {Our discussion will be agnostic as to the cause of $G\ne\GN$; in practice the precise model for $G$ would be input at a later stage when comparing the predictions made here with observations.}

\subsection{Cepheid Stars}

\begin{figure}
    \centering
    \includegraphics[width=0.45\textwidth]{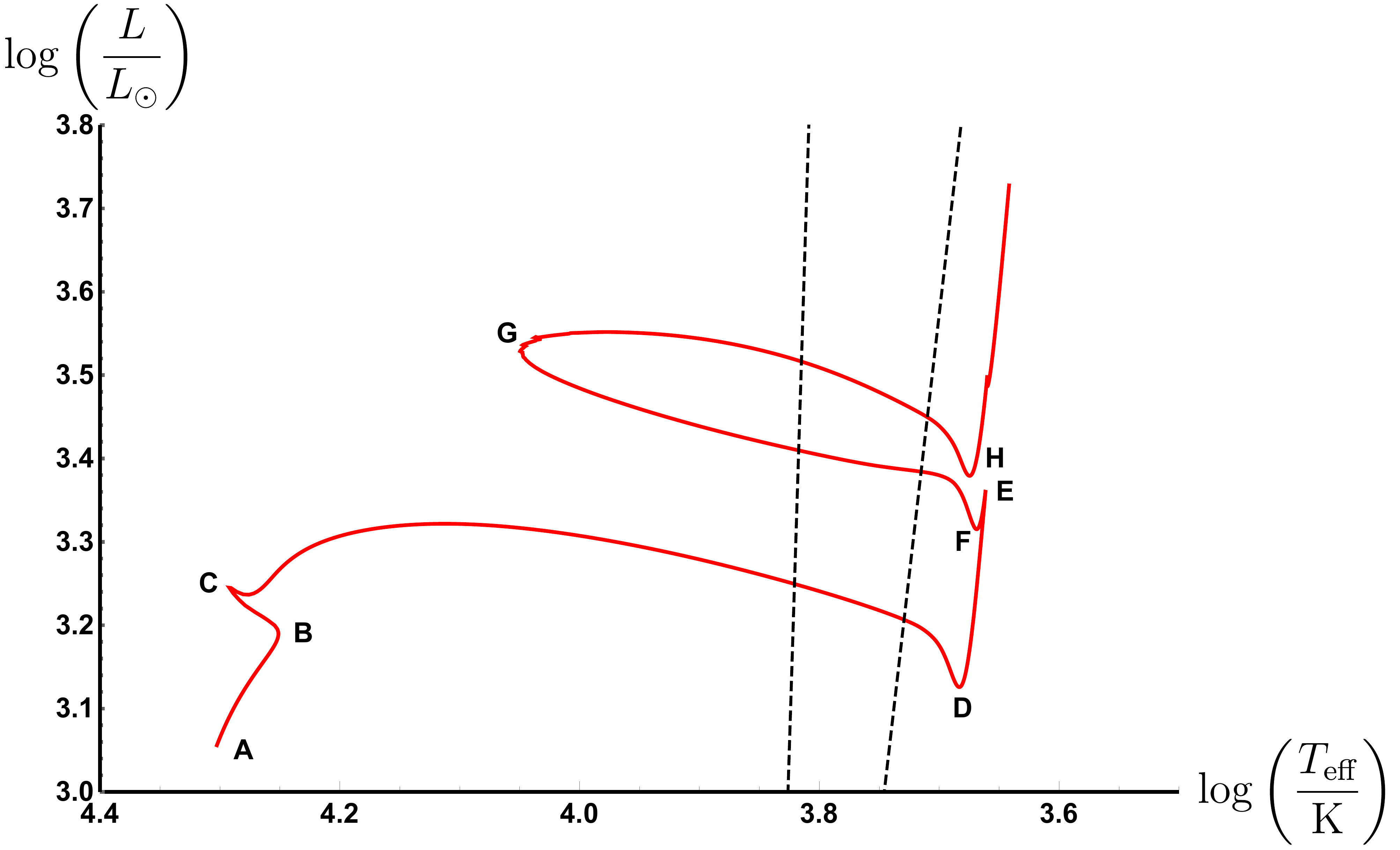}
    \caption{The evolution of a $5M_\odot$ star with $Z=0.0006$ in GR (solid, red). The dashed black lines show the edges of the instability strip. (A) ZAMS. (B) Core contraction begins. (C) Central hydrogen exhaustion. (D) Envelope becomes convective. (E) Central hydrogen burning begins. (F) Blue loop begins. (G) Furthest point on the blue loop. (H) Central helium exhausted.}
    \label{fig:5MGR}
\end{figure}

Cepheids are post-main-sequence stars of mass $\sim 3$--$20M_\odot$ that pulsate with a period that is dependent primarily upon their luminosity with a weak dependence on mass and metallicity. They are composed of a mainly inert convective helium core (there is a small amount of helium burning at the very center of the core) surrounded by a thin hydrogen burning shell that is their primary power source. Surrounding this is a convective envelope. The envelope contains a thin layer of partially-ionized helium that drives the pulsations. In particular, a small contraction of the star will cause a rise in the nuclear burning rate, which is accompanied by a temperature increase. This would typically result in an increase in the outward pressure due to photon absorption in the outer layers but in these stars it instead causes further helium ionization, which causes an increase in the opacity. This makes further ionization easier, leading to further contractions. This ionization zone therefore acts as an energy dam, storing energy that is eventually released, resulting in the outward phase of the pulsation. This opacity-driven process is known as the \emph{$\kappa$-mechanism} \cite{1980tsp..book.....C}. 

Figure \ref{fig:5MGR} shows the typical evolution of a Cepheid star from its zero age main-sequence (ZAMS) (point A) through central helium exhaustion (point H); we have chosen a 5$M_\odot$ star but any mass will exhibit qualitatively similar features. The $\kappa$-mechanism can only drive pulsations in the narrow region between the two dashed lines referred to as the \emph{instability strip}. This is because the internal stellar motion must be non-adiabatic in order for the ionization zone to store energy. If it is not then the energy losses above the ionization zone would necessarily balance the energy gains at its base. The blue edge of the instability strip corresponds to the stage of stellar evolution where the ionization zone enters the non-adiabatic region and the red edge corresponds to the stages where convective motion damps the pulsations sufficiently that they are essentially quenched. Examining the figure, one can see that a star will cross the instability strip at least three times (it is possible to have more crossings between (E) and (H)). The \emph{first crossing} occurs between phases (C) and (D) and lasts for a very short time ($\sim 10^4$ yr), making the probability of observing stars here extremely low. The \emph{second crossing} and \emph{third crossing} occur during phases (F)--(H) during the so-called \emph{blue loop}, which lasts $\sim1$--$100$ Myr, and it is on these crossings that Cepheids are typically observed.

The blue loops are the result of the interplay of several complex physical processes that are not fully understood \cite{2004A&A...418..213X,2015MNRAS.447.2951W,2019arXiv190310423S}. Indeed, Fig. \ref{fig:5MGR} shows the track for a star that has executed a blue loop but the size, shape, and even existence of the loop are highly dependent on the prescription one employs for these processes. For this reason, it will prove instructive to understand the salient features governing each phase of evolution in Fig. \ref{fig:5MGR} to aid us in constructing models that exhibit blue loops when $G(\rdm)\ne\GN$. We caution that our review will be far from comprehensive and we refer the reader to the exhaustive literature (e.g. \cite{1990sse..book.....K}) on this ongoing field of research for more in-depth discussions.

Point (A) is the ZAMS. At this point the star begins to burn hydrogen in its core, depleting its supply and producing helium. At point (B) the hydrogen core begins to contract. This is the first place where internal processes affect the post-main-sequence dynamics. The amount of contraction and overshooting will determine the final mass and radius of the core, as well as the hydrogen and helium gradients above the core. Point (C) corresponds to core hydrogen depletion. At this point, the core is composed of inert helium surrounded by a thick hydrogen burning shell. A large fraction of the energy generated by this shell is absorbed by the outer layers of the star, causing a rapid expansion and cooling of the envelope. This is the reason for the reddening of the star in the color-magnitude diagram (CMD). This phase spanning the region between (C) and (D) lasts $\sim 10^4$ yr. This is the second phase where internal processes are important. The efficiency of convective mixing and the length scale over which convective mixing occurs plays an important role in determining the temperature and luminosity of point (D). Smaller mixing lengths move it to lower temperatures. Point (D) is the Hayashi track. This is the point where the envelope has become fully convective. The core temperature during this phase can exceed the ignition temperature for the triple-alpha process so some helium burning occurs in the core but the star's luminosity is primarily due to hydrogen burning in the thin shell surrounding the core. The star is a red giant between (D) and (E). Point (E) is referred to as the \emph{dredge-up}. The convective envelope has extended into the region that was formerly the convective core during main-sequence burning so that material that has been processed through the CNO cycle is mixed throughout the entire envelope. 

The blue loop may begin at point (F), when the envelope becomes radiative. The blue-loop phase is well-studied numerically and is a generic feature of many numerical codes (e.g. \cite{2015ApJS..220...15P,Paxton:2017eie,2019arXiv190301426P}) but it is not the result of any one single simple physical process. 
Whether or not the star executes a blue loop depends on the size of the core, the amount of overshooting (overshooting increases the helium core mass at the end of hydrogen burning, outward overshooting from the core can inhibit the loops whereas inward shooting from the base can exacerbate them \cite{2004A&A...418..213X}), and the amount of convective mixing. Mass loss and rotation can also be important. One essential requirement for the execution of the loops is efficient semiconvection \cite{2019arXiv190310423S}. There are two inequivalent criteria for convection: the Schwarzschild criterion and the Ledoux criterion (see e.g. \cite{1995MNRAS.275..983E}). In homogeneous media the two are equivalent but in the presence of strong gradients in the chemical composition it is possible that the Schwarzschild criterion is violated but the Ledoux criterion is not. In such \emph{semi-convective} regions, mixing is slow. For Cepheid stars at stage (F), the amount of semiconvection determines the efficiency with which helium is mixed into the core. Point (G) is the hottest part of the blue loop and point (H) indicates the end of core helium burning where the star rejoins the Hayashi track. All of the processes described above lack a completely fundamental description and are typically implemented into stellar structure codes using phenomenological descriptions controlled by efficiency parameters that are likely mass dependent. These are used to understand the role and importance of these processes for various stages of stellar evolution. 

For stars with $3$--$9M_\odot$, stellar modelling indicates that the blue loops are abundant provided that semiconvection is included in the model. For heavier stars ($10$--$20M_\odot$), the situation is more complicated \cite{2019arXiv190310423S}. For these stars, core helium burning can begin before point $D$ in Fig. \ref{fig:5MGR}. Whether or not this happens depends strongly on the mixing processes described above and it is possible that the star can end up as a red supergiant, a blue supergiant, or spend some time in both phases (blue loops). For these stars, the existence of blue loops is highly-dependent on the choice of the efficiency parameters for the mixing processes. 

\subsection{The Period--Luminosity Relation}

When a star crosses the instability strip to become a Cepheid it obeys a well-known relationship between pulsation period, $P$, and luminosity, $L$ \cite{Freedman:2010xv}:
\begin{equation}\label{eq:PLRGR}
    \log(L)=A\log (P) + \epsilon\log(T_{\rm eff}) +\beta
\end{equation}
with $A\simeq1.3$ \cite{Becker:2015nya}. Heuristically, this relationship comes about as follows. Using stellar perturbation theory, one can show that the period of any stellar oscillation satisfies \cite{1980tsp..book.....C,Sakstein:2013pda,Sakstein:2016lyj} 
\begin{equation}\label{eq:PG}
    P\propto \sqrt{\frac{R^3}{\GN M}}.
\end{equation}
Using the Stefan-Boltzmann law, $L\propto R^2 T_{\rm eff}^4$, we can eliminate the radius in favour of the luminosity and the effective temperature, and we can further eliminate the mass $M$ by assuming a mass-luminosity relation\footnote{There is some residual mass and metallicity-dependence since a core mass-radius relation is more appropriate for shell-burning stars.}. Thus we are left with a relation between the period, the effective temperature, and the luminosity, which can always be written in the form of Eq. \eqref{eq:PLRGR}. In practice, the coefficients $A$, $\epsilon$, and $\beta$ must be fit to data rather than derived from first principles.

\begin{figure*}
    \centering
    \includegraphics[width=0.45\textwidth]{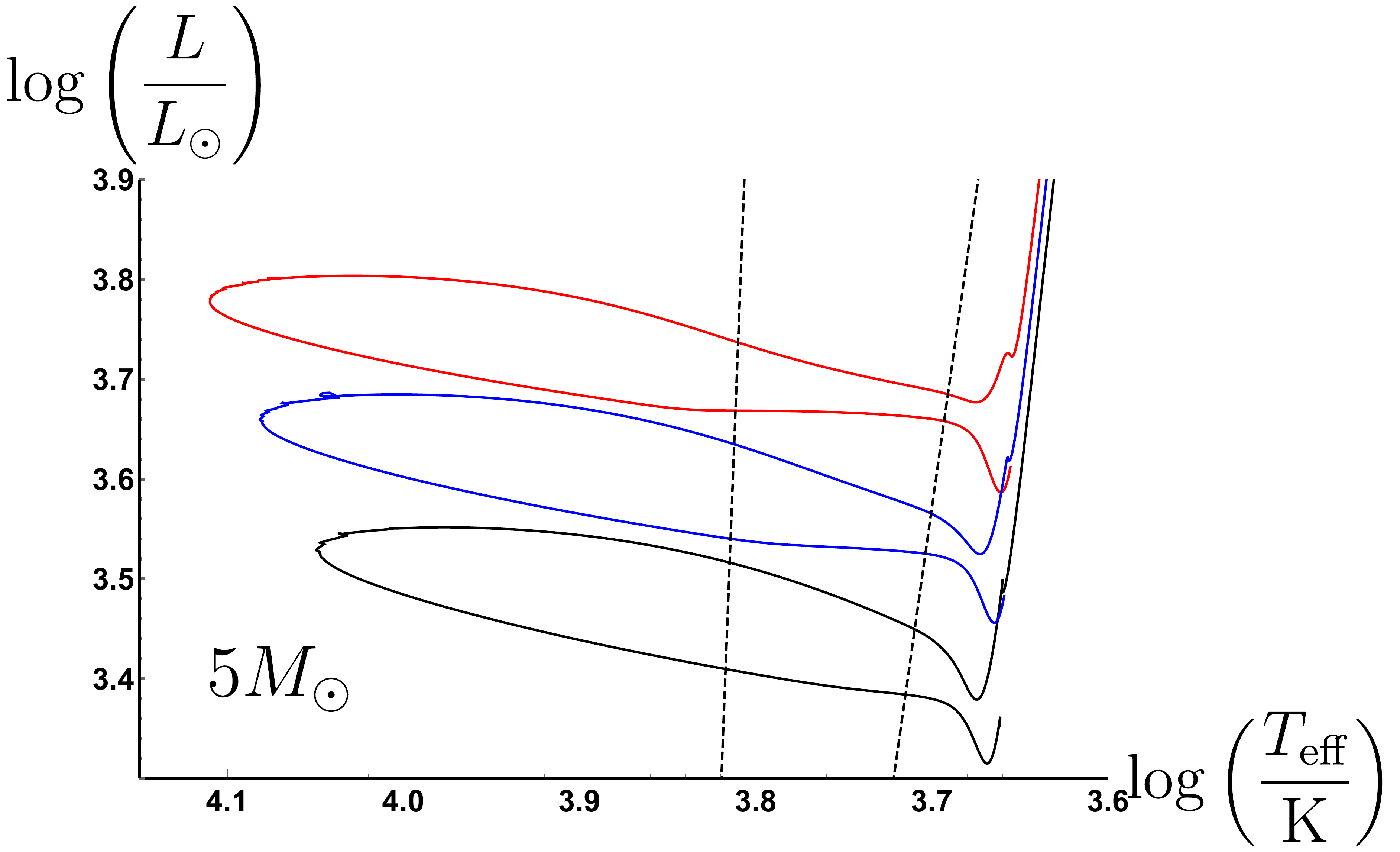}
    \includegraphics[width=0.45\textwidth]{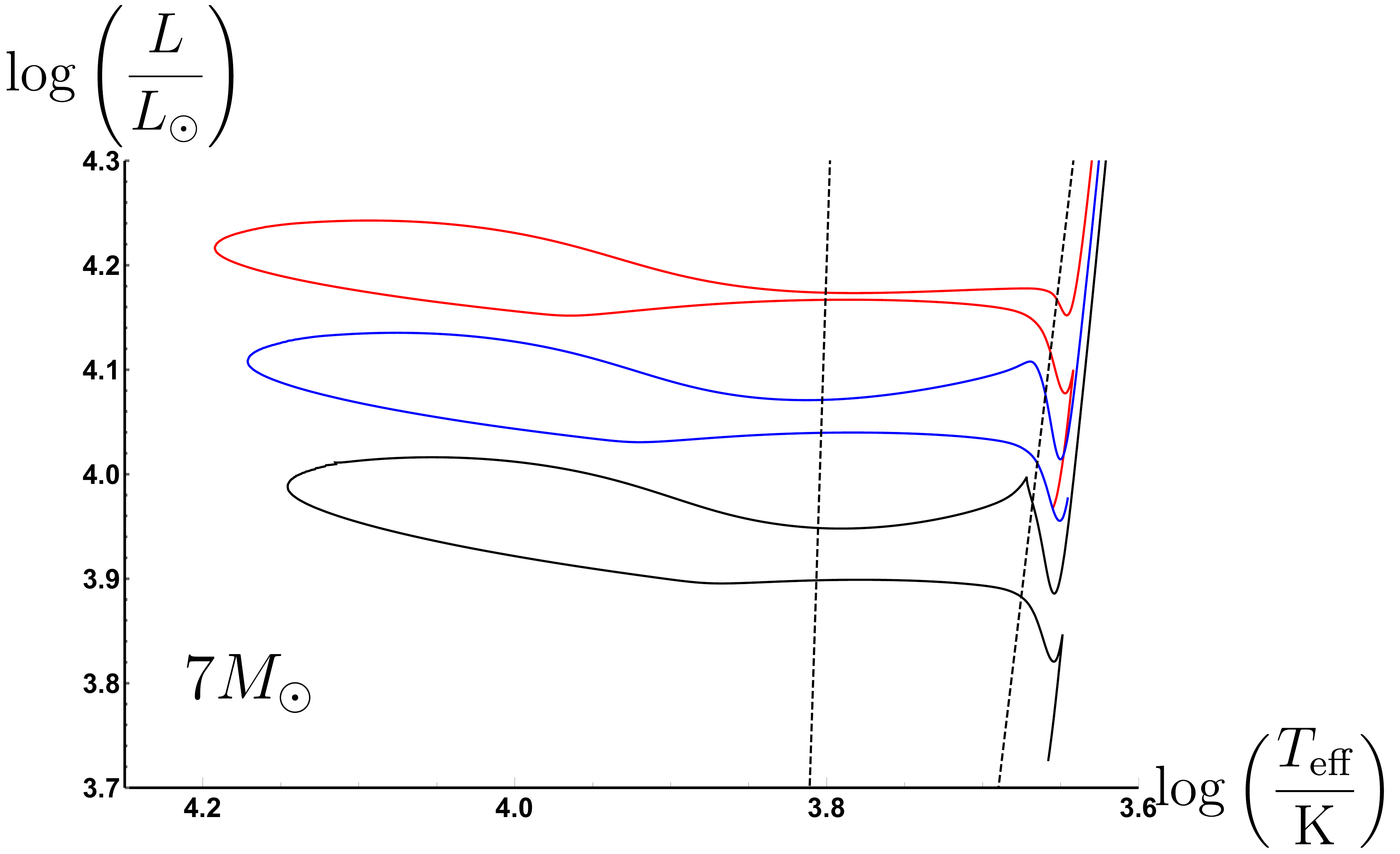}
    \includegraphics[width=0.45\textwidth]{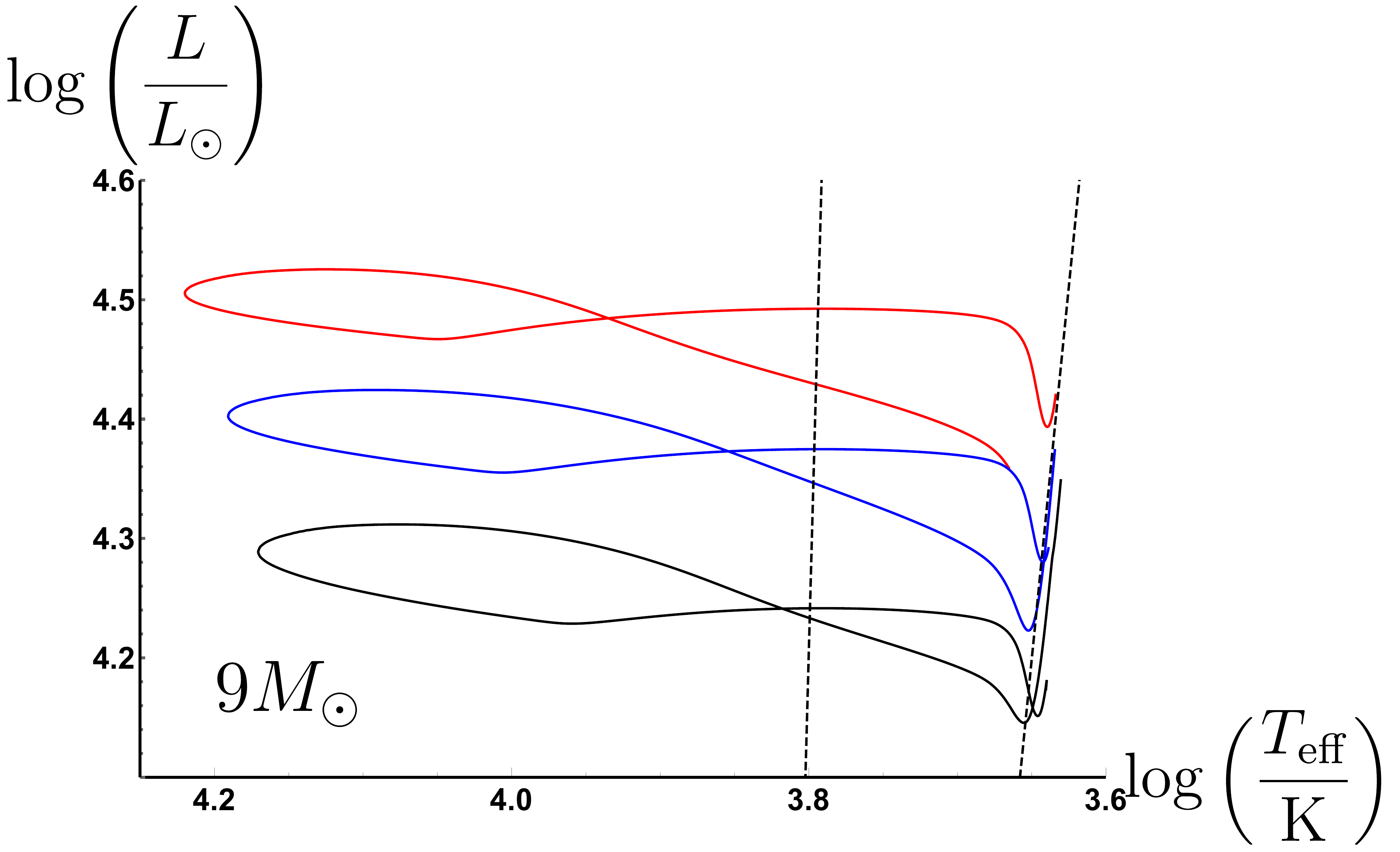}
    \includegraphics[width=0.45\textwidth]{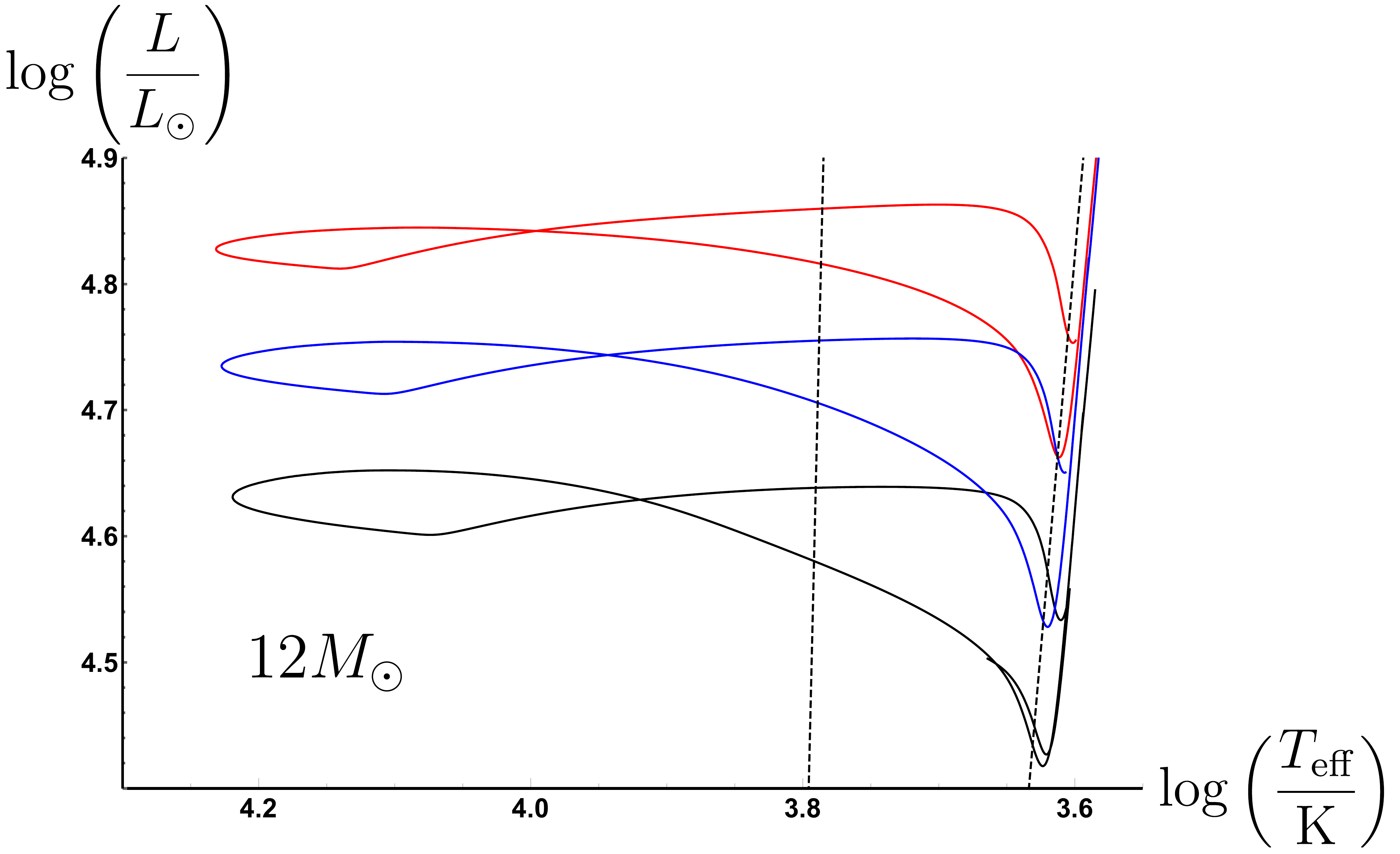}
    \caption{The blue loops for representitive stellar models used to derive the coefficients $B(M)$ given in Table \ref{tab:Bs}, which appear in the PLR in Eq. \eqref{eq:PLRUN}. The black (lower) lines show GR models ($\Delta G/\GN=0$), the blue (middle) and red (upper) lines correspond to $\Delta G/\GN=0.075$ and $\Delta G/\GN=0.15$ respectively. The black dashed lines show the edges of the instability strip. Each panel corresponds to a different mass, as indicated. Tracks for other masses are qualitatively similar.}
    \label{fig:loops1}
\end{figure*}

\subsection{The Unscreened Period--Luminosity Relation}

We now wish to understand how the PLR changes in galaxies less dense than the MW that may be unscreened so that $G(\rdm)>\GN$. Examining Eqs. \eqref{eq:PLRGR} and \eqref{eq:PG}, one can discern two potential effects. First, since $P\propto G(\rdm)^{-1/2}$, increasing $G(\rdm)>\GN$ reduces the period, i.e. the star pulsates faster. The second effect is that we expect the luminosity at fixed mass to be larger. It is well-documented that increasing the strength of gravity makes stars more luminous at fixed mass due to an increase in the nuclear burning rate that is necessary to produce the pressure gradient needed to sustain hydrostatic equilibrium (see \cite{Sakstein:2018fwz} and references therein). The change to the PLR can be written as
\begin{equation}\label{eq:PLRUN}
    \Delta \log(L)=\frac{A}{2}\log\left(1+\frac{\Delta G}{\GN}\right)+B(M)\log\left(1+\frac{\Delta G}{\GN}\right),
\end{equation}
where $\Delta \log(L)\equiv \log(L)-\log(L_\text{GR})$ is the difference between the true luminosity and that which would be inferred if one had used the GR relation given in Eq. \eqref{eq:PLRGR}. The term proportional to $A$ is the effect of changing the pulsation period (see Eq. \eqref{eq:PG}) whereas the second is an ansatz parameterizing the change in the luminosity due to the increased rate of nuclear burning. We will justify this \emph{post hoc} by showing that a power-law relation is a good approximation. It is possible that the free coefficient $B$ is a function of the stellar mass so we have included this possibility. We will also find the value of $B$ to depend on whether the Cepheid is observed at the second or third crossing of the instability strip.

\begin{table}[ht]
  \begin{center}
    \begin{tabular}{|c|c|c|}
      \hline
       $M/M_\odot$ & $B$ at second crossing & $B$ at third crossing\\ 
      \hline
    \rule{0pt}{3.5ex}
      5 & 4.21 & 3.67\\
    \rule{0pt}{3.5ex}
      6 & 4.52 & 3.61\\
    \rule{0pt}{3.5ex}
      7 & 4.45 & 3.79\\
    \rule{0pt}{3.5ex}
      8 & 4.34 & 3.58\\
    \rule{0pt}{3.5ex}
      9 & 4.18 & 3.46\\
      \rule{0pt}{3.5ex}
      10 & 4.00 & 3.48\\
      \rule{0pt}{3.5ex}
      11 & 3.81 & 3.58\\
      \rule{0pt}{3.5ex}
      12 & 3.67 & 3.92\\
      \rule{0pt}{3.5ex}
      13 & 3.58 & 3.95\\
      \hline
    \end{tabular}
  \caption{Slope $B(M)$ of the $\Delta \log(L) - \log(1+\Delta G/G_{\rm N})$ relation given in Eq. \eqref{eq:PLRUN} for a range of Cepheid masses measured at second or third crossing of the instability strip. }
  \label{tab:Bs}
  \end{center}
\end{table}

In order to calculate $B$ we have modified the stellar structure code MESA \cite{2011ApJS..192....3P,2015ApJS..220...15P,Paxton:2017eie,2019arXiv190301426P} to compute stellar evolution for general $G(\rho_\text{DM})$. We have evolved a grid of $5M_\odot$--$13M_\odot$ models with $Z=0.0006$ from the pre-main-sequence through the blue loop phase in order to calculate $\Delta\log(L)$. We varied $\Delta G/\GN$ in the range $0\le \Delta G/\GN\le 0.15$. For masses $3M_\odot<M<9M_\odot$ we found the loops to be generic provided that semiconvection and the Ledoux criterion are enabled. This allowed us to use a fixed set of mixing parameters. We chose the mixing length parameter to be $\alpha_{\rm MLT}=1.73$ ($\alpha_{\rm MLT}\equiv \lambda_{\rm MLT}/H_P$ where $\lambda_{\rm MLT}$ is the mixing length and $H_P$ is the pressure scale height). The efficiency of semiconvection was taken to be $\alpha_{SC}=0.1$. For overshooting, we adopted the parameters $f=0.014$ and $f_0=0.004$ for burning and non-burning cores of all compositions. See the instrumentation papers \cite{2011ApJS..192....3P,2015ApJS..220...15P,Paxton:2017eie,2019arXiv190301426P} and \cite{2000A&A...360..952H} for the details of the overshooting parameters. For masses in the range $10M_\odot$--$13M_\odot$ we found the loops to be uncommon (as expected \cite{2019arXiv190310423S}) and it was necessary to vary the mixing parameters as a function of $\Delta G$ in order to produce models that exhibited them. We used $1.1\le\alpha_{\rm MLT}<1.3$ and $10^{-4}<\alpha_{\rm SC}<100$ in order to achieve this. Overshooting was as above. Different parameter choices lead to small variations in the blue-loop shape and luminosity at fixed mass, which is part of the scatter in the PLR that is present even in GR. Ideally this uncertainty would be included in any statistical analysis that applies the PLR we derive here to find observational bounds on $\Delta G$. 

\begin{figure*}[ht]
    \centering
    \includegraphics[width=0.45\textwidth]{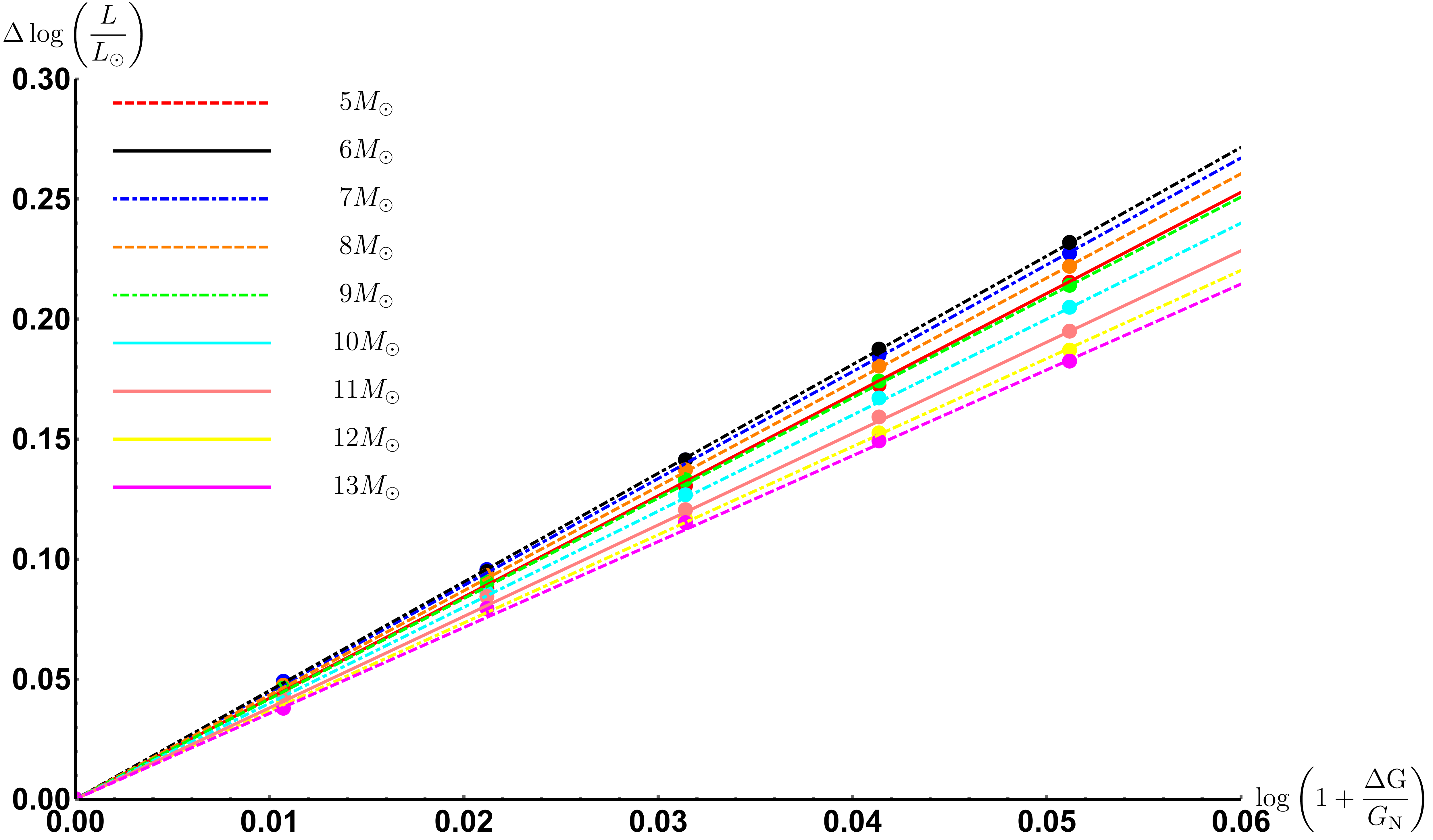}\hfill
    \includegraphics[width=0.45\textwidth]{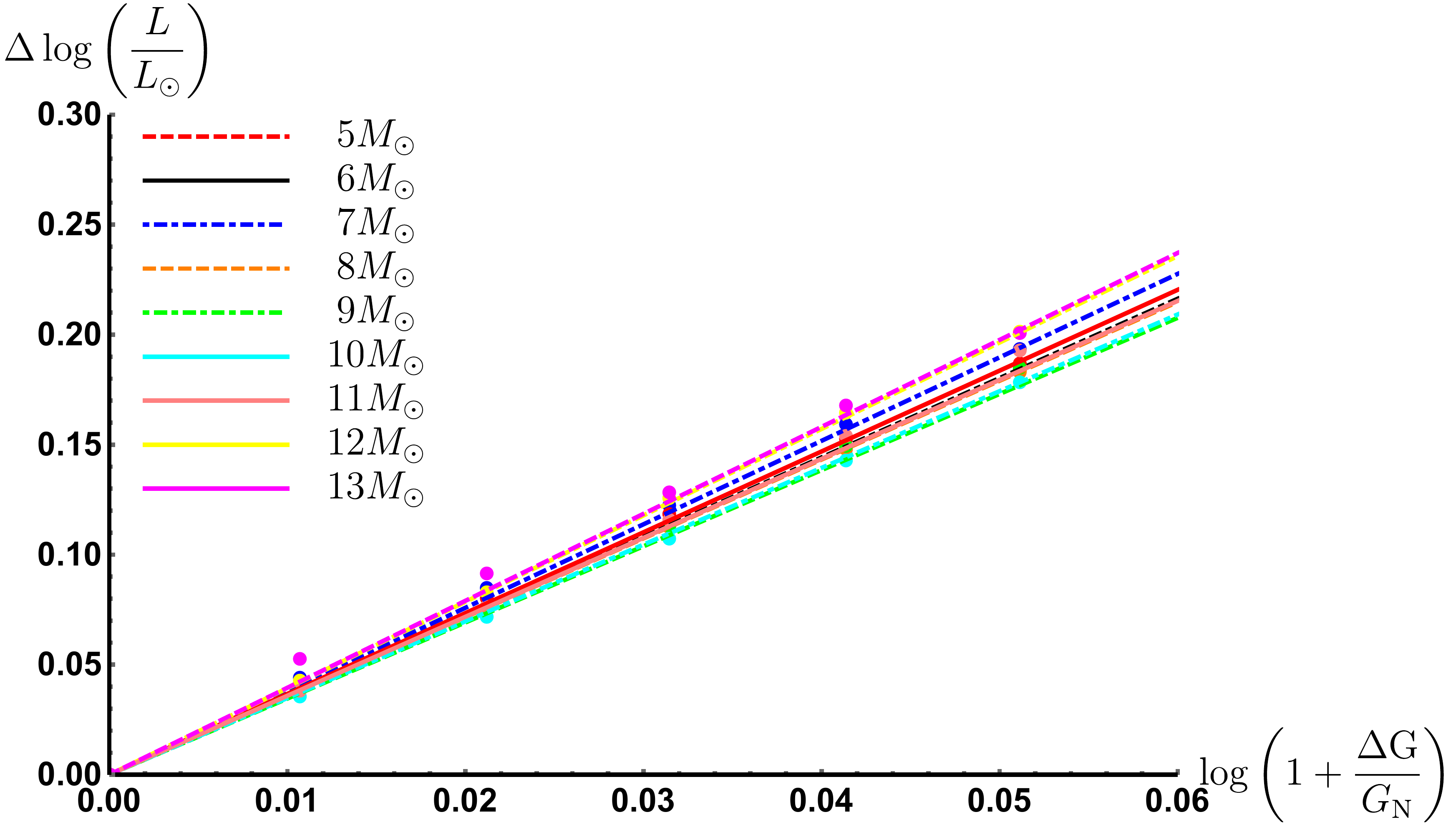}
    \caption{$\Delta\log(L)$ as a function of $\log(1+\Delta G/\GN)$ at the blue edge of the instability strip for the second crossing (left) and the third crossing (right). The points indicate the results of MESA models and the lines are the best-fit linear relations, the slopes of which are the coefficients $B(M)$ given in Table \ref{tab:Bs}. }
    \label{fig:deltaL}
\end{figure*}

The resultant loops for some representative masses are shown in Fig. \ref{fig:loops1} (for visual clarity we do not show the pre-loop phase). The loops for other masses not shown are qualitatively similar. We have calculated the coefficients $B(M)$ given in Eq. \eqref{eq:PLRUN} at each crossing of the blue edge of the instability strip (this is a convenient comparison point and the physics is better-understood here) by performing a linear fit for each mass. The coefficients are given in Table \ref{tab:Bs} and visual representations of our models and the best-fitting relations are given in Fig. \ref{fig:deltaL}. Future observational tests of this screening mechanism using unscreened galaxies (with dark matter densities smaller than the MW) can utilize these coefficients to make the necessary theoretical predictions to constrain the theory. 

\section{Main-Sequence Stars}
\label{sec:MS}

Main-sequence stars have been less successful at constraining screening mechanisms than post-main-sequence stars, primarily because they are subject to degeneracies that are more difficult to break. For example, fifth force effects such as luminosity enhancements are degenerate with the metallicity \cite{Koyama:2015oma}.\footnote{The one exception to all this is the Sun, where we have precision measurements of quantities such as the solar neutrino flux and access to thousands of oscillation modes. Constraints from the Sun at the level of $\Delta G/\GN\sim 10^{-2}$ have been obtained by considering the effects on the seismic solar model \cite{2003MNRAS.341..721L}, but these do not apply to theories where the Solar System is screened.}  One promising test is that they may produce novel deviations in the galactic properties as a whole as a result of their integrated effects \cite{Davis:2011qf}, so we briefly study their properties here. We will do this both analytically and numerically (using MESA). Both approaches have previously been used in the literature to derive the properties of main-sequence stars in other screening models \cite{Davis:2011qf,Koyama:2015oma}, and Ref. \cite{Adams:2008ad} has studied the effects of changing the fundamental constants on some stellar properties.


\subsection{Analytic Expectations}
\label{sec:ESM}

The Eddington standard model is a simple analytic model for main-sequence stars. It is certainly not detailed enough to produce accurate models for individual objects, but it has proved useful in the study of screened stars for discerning the main effects of changing fundamental physics on the gross stellar properties \cite{Davis:2011qf,Koyama:2015oma}. This is precisely because the lack of detailed microphysics and time-evolution allows one to isolate the effects of changing $G$. Its use in understanding how screening mechanisms affect main-sequence stars is standard in the literature \cite{Davis:2011qf,Koyama:2015vza} so we will not reproduce its derivation in detail, but we do provide a brief review for the unfamiliar reader in Appendix \ref{sec:poly}. There we introduce polytropic stellar models and derive the formulae we use in this section.

The main result of the Eddington standard model is an analytic expression for the luminosity as a function of mass (Eq. \eqref{eq:LEDD})
\begin{equation}\label{eq:L(M)}
    L=\frac{4\pi (1-\beta(M))\GN M}{\kappa},
\end{equation}
where $\kappa$ is the (assumed constant) opacity, and $\beta (M)$ is the ratio of the gas pressure to the total pressure (the ratio of the radiation pressure to the total pressure is $1-\beta(M)$). $\beta(M)$ is found by solving a quartic equation given in Eq. \eqref{eq:Eddquartic}. Stars that are entirely gas supported have $\beta=1$ so that there is no luminosity\footnote{This is a drawback of the approximations made in Eddington standard model. Setting $\beta=1$ is tantamount to removing photons from the star entirely.} and stars which have $\beta=0$ are entirely radiation supported and the luminosity is maximum (the Eddington luminosity). A careful analysis of Eq. \eqref{eq:Eddquartic} reveals that $\beta(M)$ is a decreasing function of mass so that low mass stars are primarily gas-supported whereas high-mass stars are primarily radiation supported. One can use simple scaling arguments to show that the luminosity of gas-supported stars scales as $G^4 M^3$ and the luminosity of radiation-supported stars scales like $G M$  \cite{Davis:2011qf,Koyama:2015vza}. In the context of screening mechanisms, this implies that low-mass stars are more sensitive probes of fifth forces than high mass stars.

\begin{figure}[b]
    \centering
    \includegraphics[width=\columnwidth]{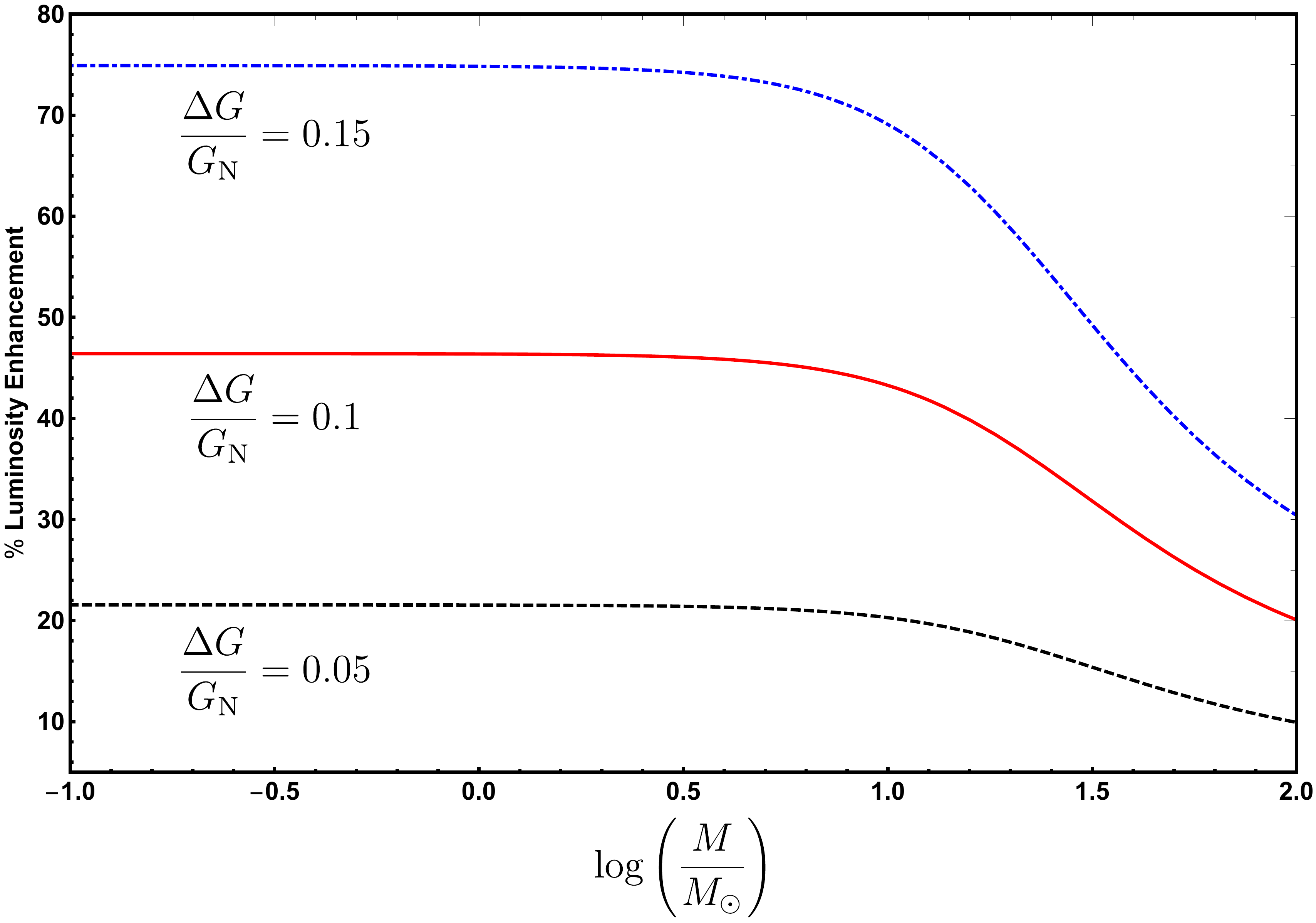}
    \caption{Analytical predictions (computed using the Eddington standard model) for the luminosity enhancement of unscreened main-sequence stars compared with screened stars of identical masses. The values of $\Delta G/\GN$ are indicated below the corresponding curves. }
    \label{fig:MSA}
\end{figure}

To investigate this quantitatively, we simply need to change $\GN\rightarrow \GN(1+\Delta G/G)$ in Eq. \eqref{eq:L(M)}.\footnote{A similar analysis to the one performed here first appeared in the following unpublished lecture notes \href{http://www.jeremysakstein.com/astro_grav_2.pdf}{http://www.jeremysakstein.com/astro\_grav\_2.pdf} by one of the authors.} There are two places where this replacement must be made: in Eq. \eqref{eq:L(M)} directly, and in Eddington's quartic equation \eqref{eq:Eddquartic} (see Eq. \eqref{eq:MEDD}). The second replacement results in the following quartic equation
\begin{equation}
    \label{eq:quarticmod}
    \frac{1-\beta(M,\Delta G)}{\beta^4(M,\Delta G)}=\left(1+\frac{\Delta G}{\GN}\right)^3\left(\frac{M}{M_{\rm Edd}}\right)^2,
\end{equation}
where the Eddington mass $M_{\rm edd}\approx 18.2\mu^{-2} M_\odot$ ($\mu$ is the mean molecular weight) is defined in Eq. \eqref{eq:MEDD}. The ratio of the luminosity of unscreened and screened stars is then
\begin{equation}\label{eq:L(M)MG}
    \frac{L_{\rm unscreened}}{L_{\rm screened}}=\left(1+\frac{\Delta G}{\GN}\right)\frac{1-\beta(M,\Delta G)}{1-\beta(M,0)}.
\end{equation}
We are interested in fifth force models where $\Delta G/\GN>0$. We therefore expect that unscreened stars are more luminous than screened stars at fixed mass since these stars require a faster nuclear burning rate in order to maintain hydrostatic equilibrium given the greater inward gravitational force. More photons are therefore produced as a byproduct. We have solved Eq. \eqref{eq:quarticmod} numerically for $\Delta G/\GN$ in the range $0$--$0.15$ and have used the resulting function to calculate the luminosity enhancement using Eq. \eqref{eq:L(M)MG}. Our results are plotted in Fig. \ref{fig:MSA}, where it is evident that unscreening main-sequences stars indeed enhances the luminosity. Furthermore, in line with our discussion above, the enhancement is indeed larger in low mass stars --- tending to the factor of $(1+\Delta G/\GN)^4$ that we predicted above --- since these are gas pressure-supported and therefore more sensitive to $G$. The factor of $\sim 75\%$ for $\Delta G/\GN\sim 0.15$ is likely a gross overestimate given that the many assumptions underlying this model greatly simplify the internal dynamics of the star and neglect many important physical processes. This is why it is of paramount importance to use numerical simulations to account for these deficiencies, which we do in the next subsection. 

\subsection{Numerical Models}

\begin{figure*}[ht]
    \centering
    \includegraphics[width=0.43\textwidth]{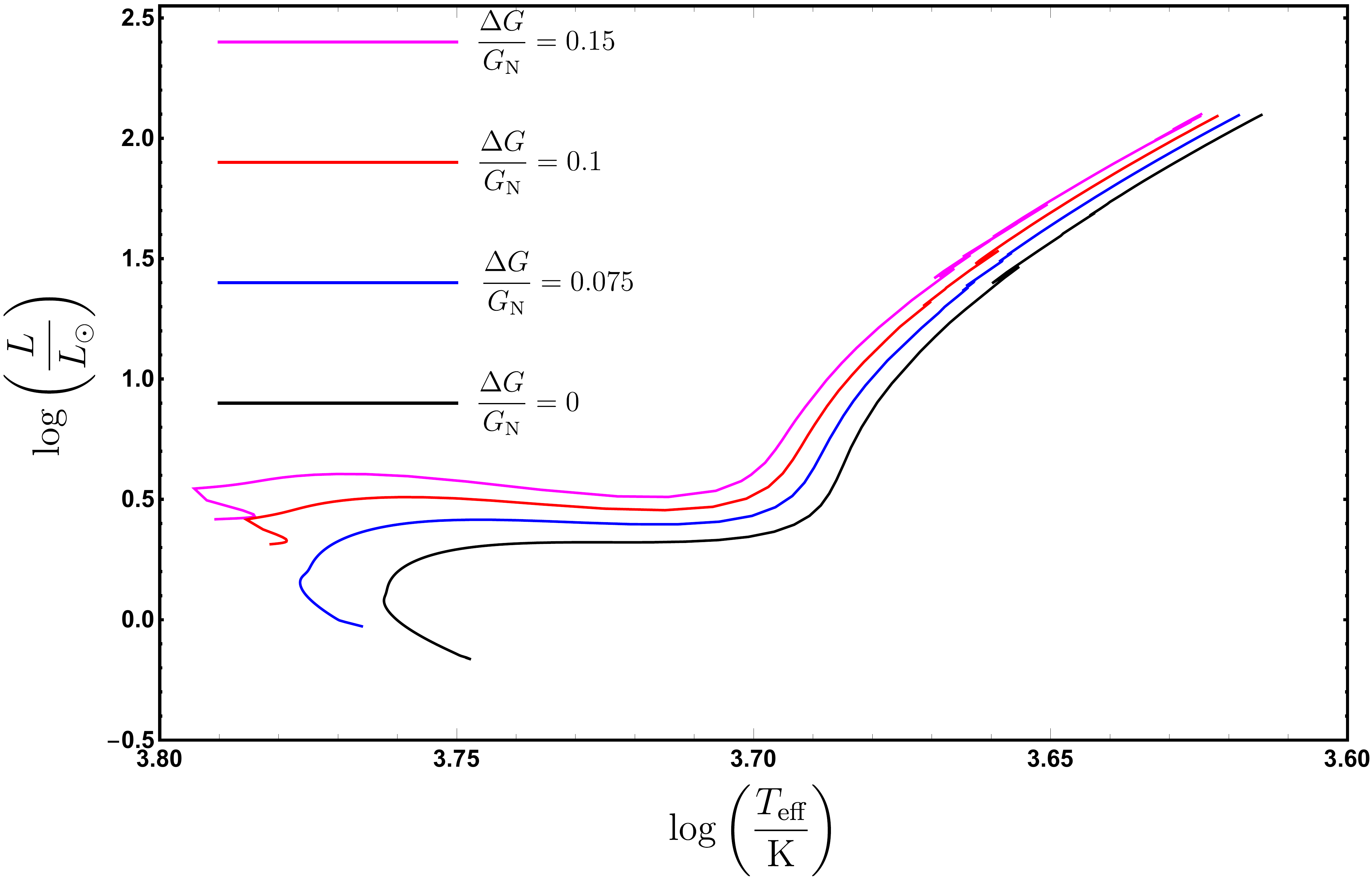}
    \includegraphics[width=0.43\textwidth]{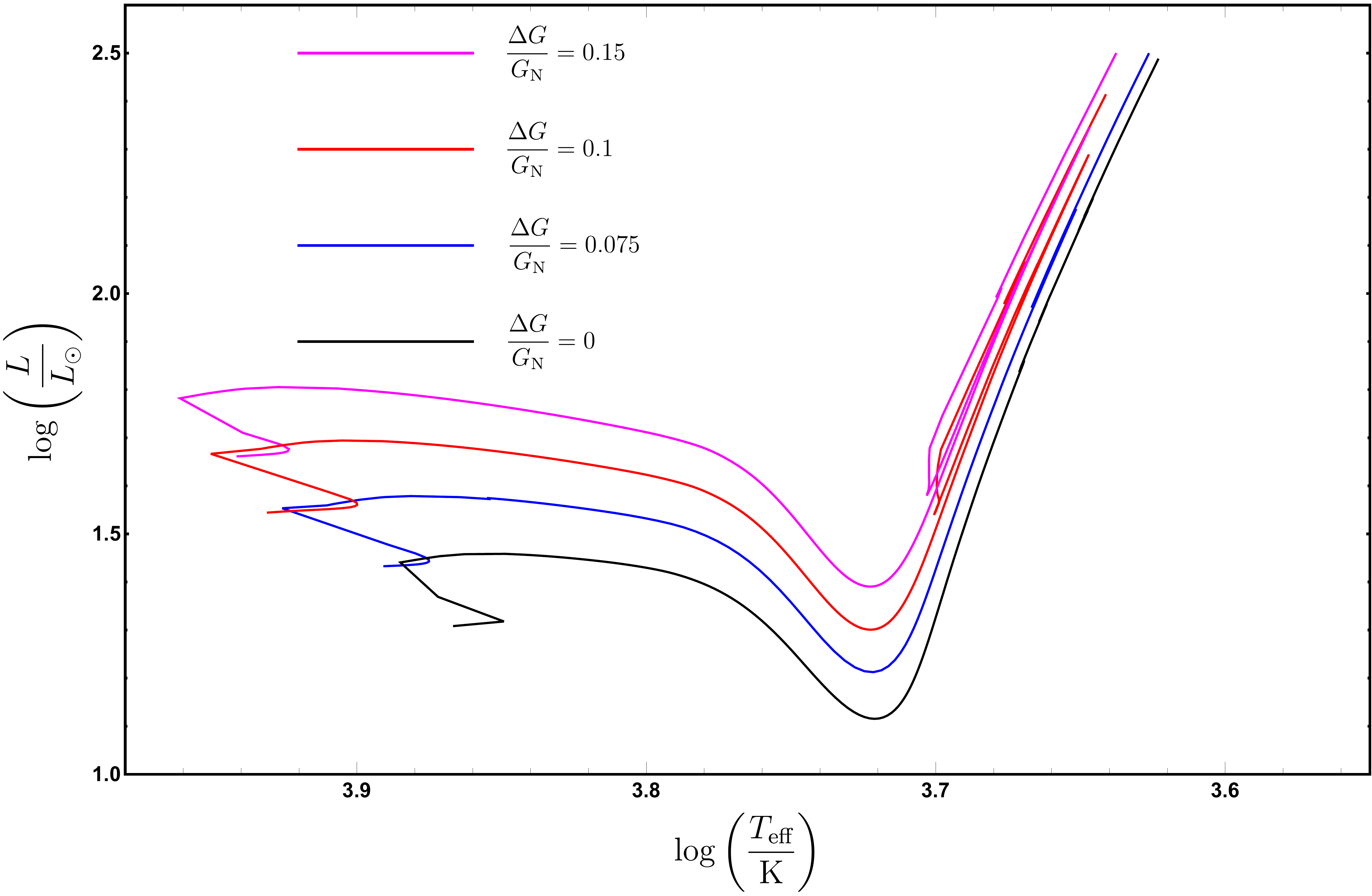}
    \caption{Main-sequence evolution for a $1M_\odot$ (left) and $2M_\odot$ (right) star of solar metallicity ($Z=0.02$) for values of $\Delta G/\GN$ as indicated in the figure.}
    \label{fig:MS}
\end{figure*}

In order to verify our expectations above, we have evolved a grid of solar mass and two solar mass models with solar metallicity ($Z=0.02$) for $\Delta G/\GN$ in the range $0\le\Delta G/\GN\le 0.15$. The evolutionary tracks in the CMD are shown in Fig. \ref{fig:MS}. Evidently, the main-sequence is indeed increasingly luminous (at fixed evolutionary point) when $\Delta G/\GN$ is increased, as is the red giant phase. In some cases, increasing $\Delta G/\GN$ can make the tracks appear to mimic those that would be exhibited by larger mass objects when $\Delta G/\GN=0$ (i.e. GR). For example, comparing the track for $\Delta G/\GN=0.15$ in the left panel with the track for $\Delta G/\GN=0$ in the right, its shape is more akin to that of a $2M_\odot$ star than a solar mass object. (The star in the left panel is cooler and less luminous than a $2M_\odot$ star so the exact object it is mimicking in GR would have a mass smaller than $2M_\odot$.) {This makes modified gravity partially degenerate with the initial mass function (IMF) in setting the overall photometric properties of galaxies.} 

\begin{figure*}[ht]
    \centering
    \includegraphics[width=0.45\textwidth]{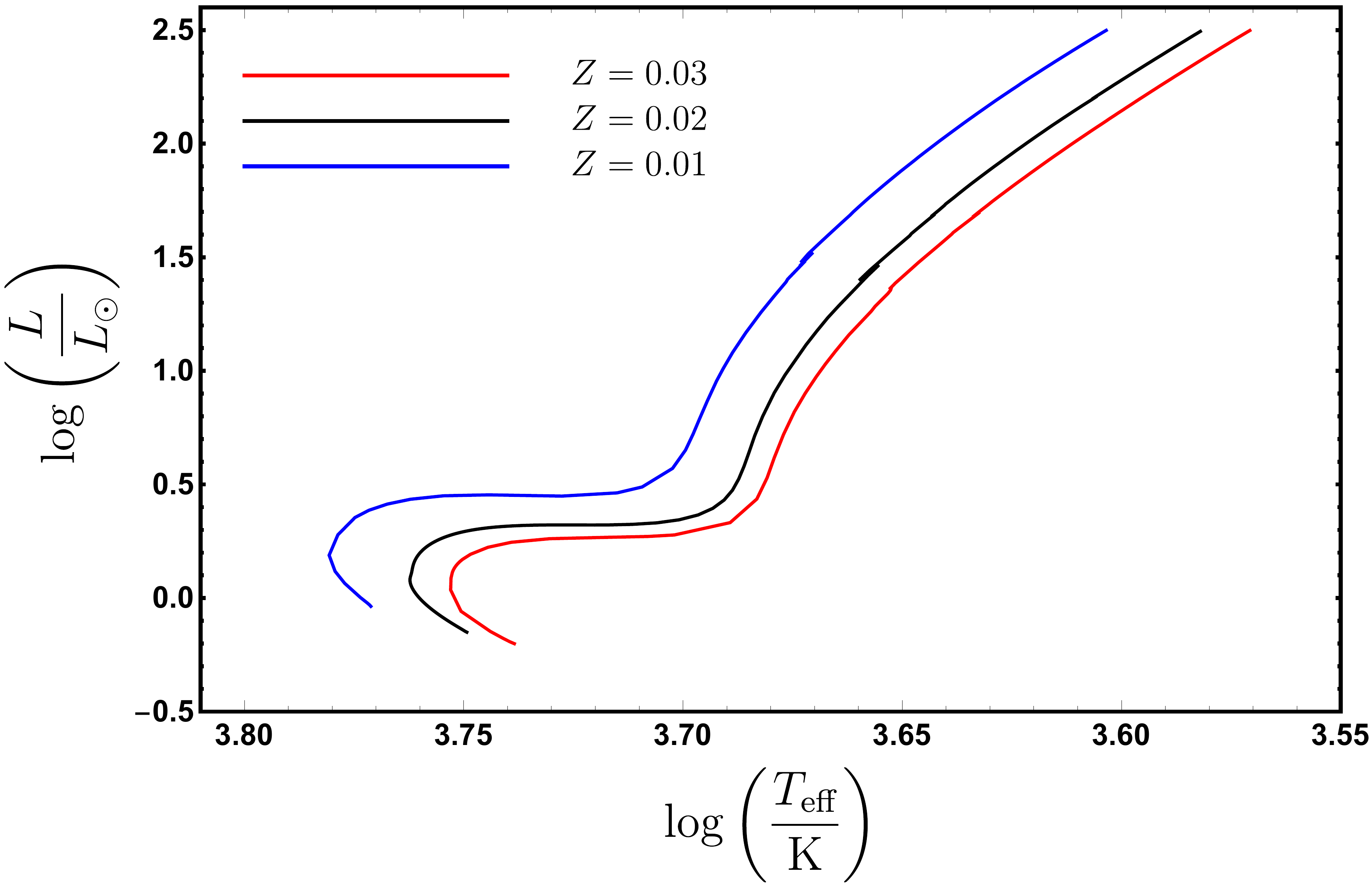}
    \includegraphics[width=0.45\textwidth]{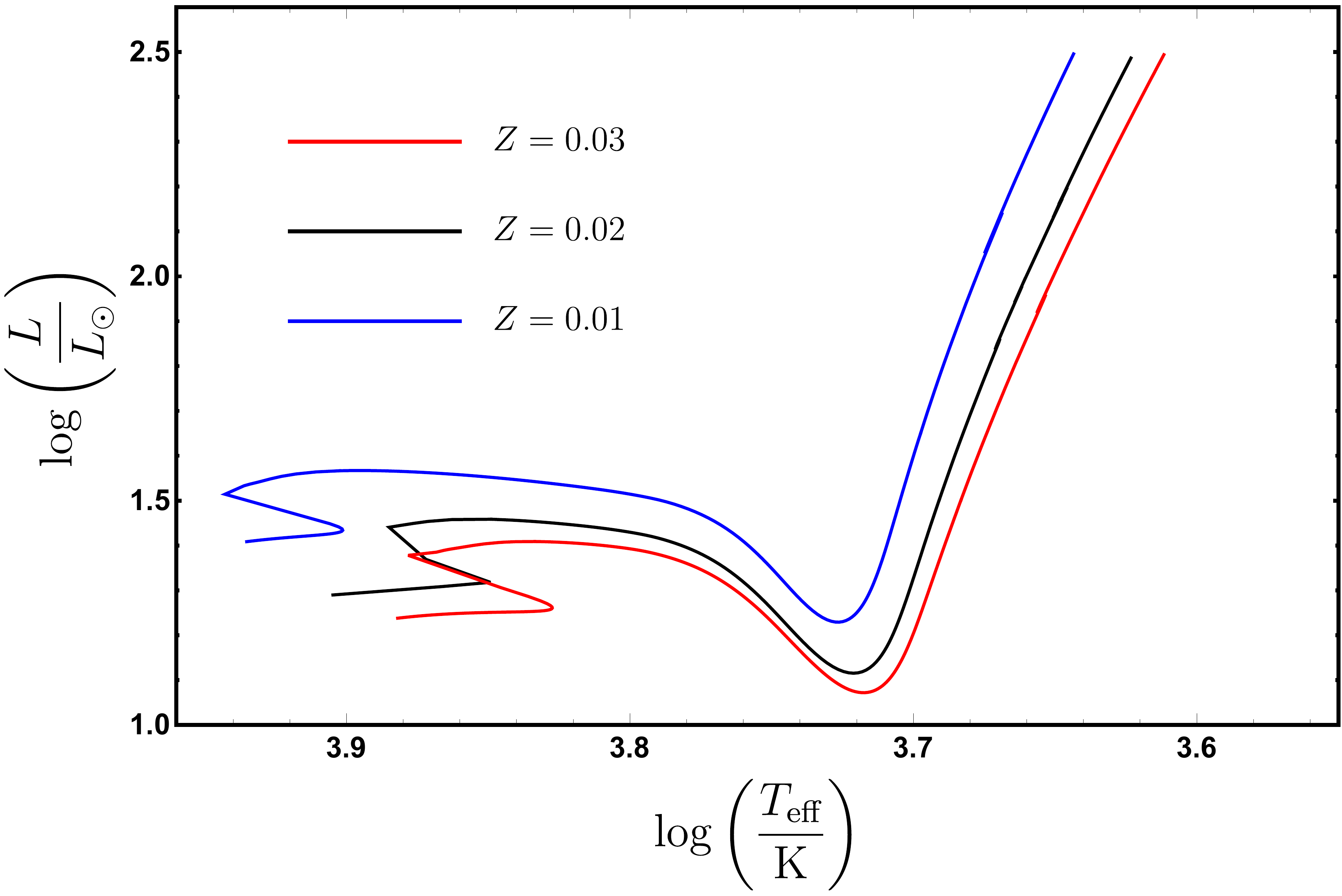}
    \caption{The evolution of solar mass (left) and $2M_\odot$ (right) stars when $\Delta G/\GN=0$ but the metallicity $Z$ is varied as indicated in the figure.}
    \label{fig:MSZ}
\end{figure*}

Unlike the Cepheid PLR, which has only a small scatter from metallicity effects, the main-sequence luminosity is very sensitive to the metallicity. This is one reason that main-sequence stars have been less successful at constraining theories that exhibit screening mechanism than post-main-sequence and dwarf stars. Any observational test based on the results derived here must mitigate this degeneracy either by measuring the metallicity or by marginalizing over it with a suitable prior. To illustrate this degeneracy, we plot the evolutionary tracks in the CMD for one and two solar mass stars in GR with $Z=0.01$ (subsolar) and $Z=0.03$ (supersolar) metallicity in Fig. \ref{fig:MSZ}. It is clear that metallicity is an important degeneracy for the main-sequence.

\section{Post-Main-Sequence Stars and the Tip of the Red Giant Branch}
\label{sec:TRGB}

One powerful method for constraining theories where the strength of gravity varies between galaxies is to compare different distance estimates to the same galaxy \cite{Jain:2012tn}. Since different indicators are sensitive to $G$ to varying degrees, the distances will only agree if one has the correct theory of gravity. We have already derived the change in the Cepheid PLR so in this section we will study another distance indicator that has previously been used to constrain modified gravity theories: the tip of the red giant branch. 

When stars of mass $0.9\lsim M/M_\odot\lsim 2$ exhaust their central hydrogen, they leave the main-sequence and evolve along the red giant branch. At this point, they are composed of a degenerate isothermal $^4$He core surrounded by a thin shell of hydrogen in hydrostatic equilibrium. Hydrogen fusion in this shell is solely responsible for the star's luminosity. As the Helium core contracts, both the core's and the shell's temperature increase. When the temperature exceeds that necessary to ignite helium burning, the triple-$\alpha$ process begins and the star moves onto the asymptotic giant branch (AGB) in a very short time. This leaves a visible discontinuity in the I-band magnitude at $I=4\pm0.1$ with a very small scatter due to metallicity effects \cite{Freedman:2010xv}. By looking for this discontinuity, the distance to the star can be calculated since both the magnitude and flux are known.

\begin{figure*}[t]
    \centering
    \includegraphics[width=0.45\textwidth]{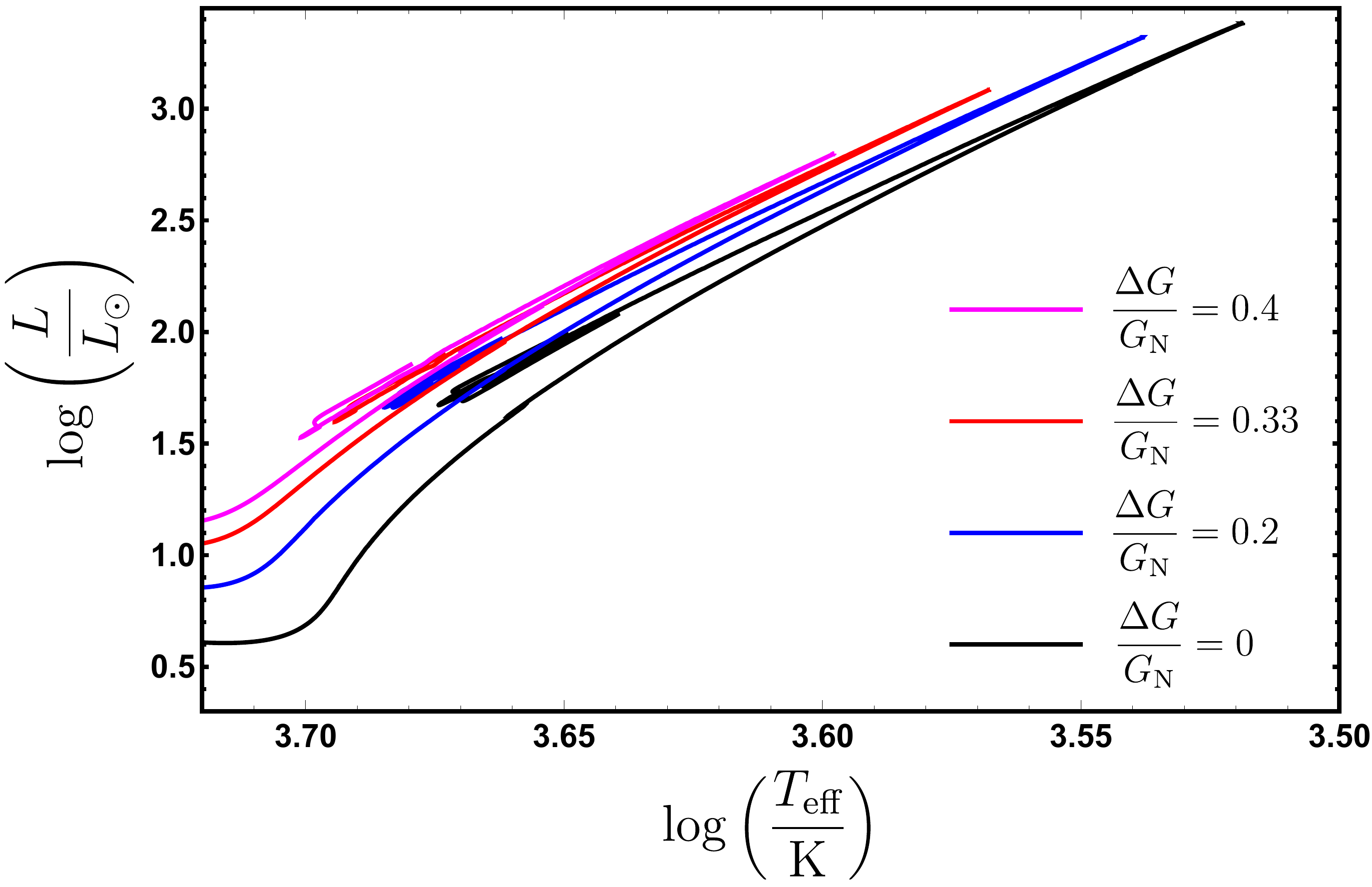}
    \includegraphics[width=0.45\textwidth]{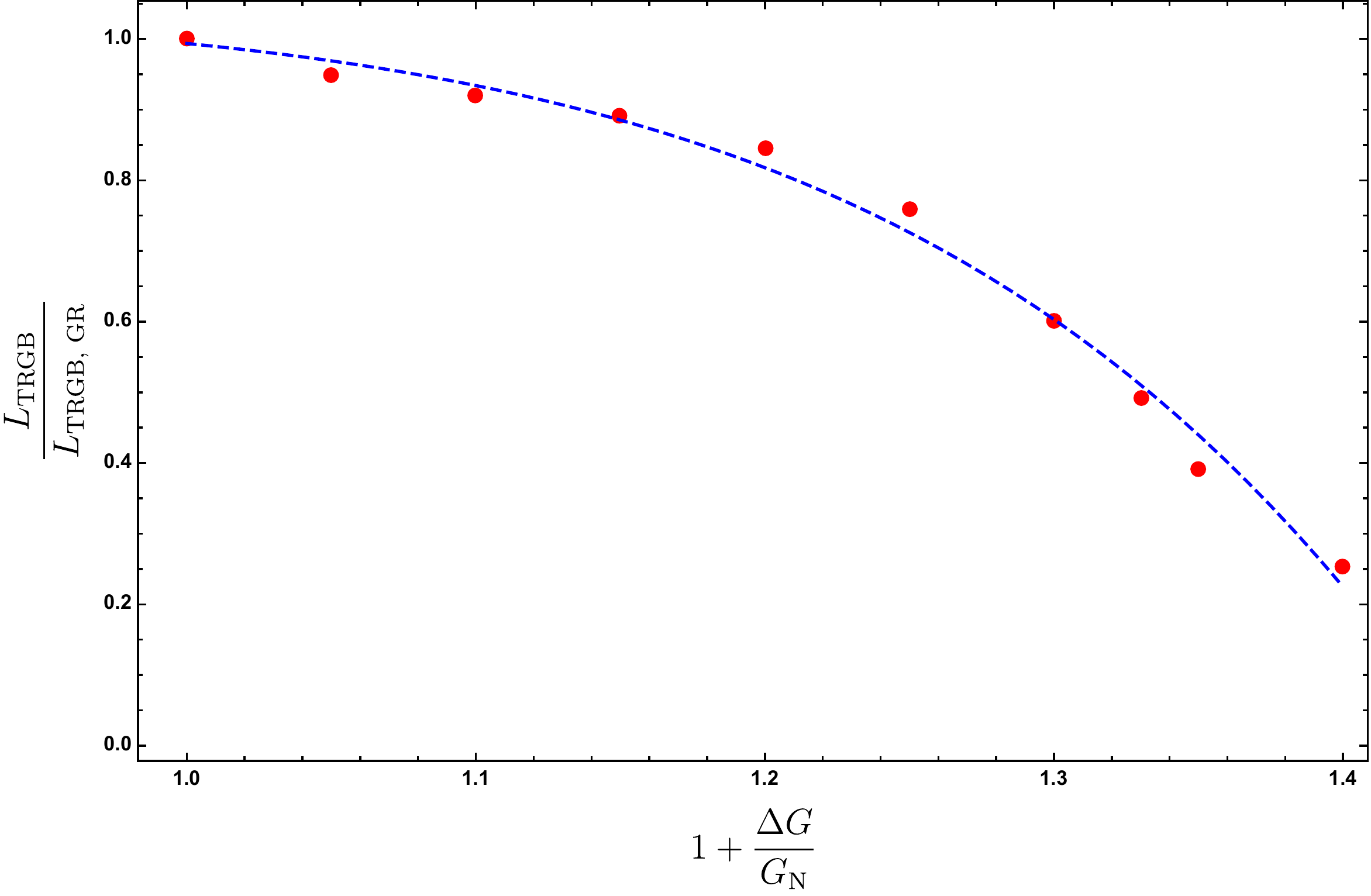}
    \caption{\emph{Left}: The Hertzsprung--Russell diagram for $1.3M_\odot$, $Z=0.02$ stars for values of $\Delta G/\GN$ indicated in the figure. Each curve terminates at the TRGB.
    \emph{Right}: The luminosity of the TRGB compared with the GR value, as a function of $1+\Delta G/\GN$. The red points are numerical MESA models and the blue dashed line is the fitting function of Eq.~\eqref{eq:LTRGB}.}
    \label{fig:TRGB}
\end{figure*}

The effects of changing $G$ on the TRGB were investigated by \cite{Jain:2012tn}. Increasing $G$ has the effect of increasing the rate of hydrogen shell burning around the core since a faster rate is needed to balance the stronger gravity. This has the effect of causing the shell and core temperature to rise at a faster rate, igniting the triple-$\alpha$ process earlier, which causes the luminosity at the tip to decrease. Using MESA, we have investigated this effect for our screening mechanism by evolving $1.3M_\odot$ stars of solar metallicity ($Z=0.02$) to the TRGB. Our results are plotted in figure \ref{fig:TRGB}. The left panel shows the Hertzsprung--Russell diagram for some representative values of $\Delta G/\GN$. One can see that low values produce similar tip luminosities whereas stronger modifications produce drastic reductions. This is quantified in the right panel where we plot the ratio of the tip luminosity to the GR value as a function of $\Delta G/\GN$. This is well-fit by the relation
\begin{equation}\label{eq:LTRGB}
    \frac{L_{\rm TRGB}}{L_{\rm TRGB,\, GR}}=1.04227\left[1- 0.0466349\left(1+\frac{\Delta G}{\GN}\right)^{8.3889}\right].
\end{equation}
Using the equation relating flux and luminosity distance, $F=L/d_L^2$, one can see that the lower tip luminosity for unscreened stars will result in the GR formula overestimating the distance:
\begin{equation}\label{eq:dTRGB}
    d_L=d_L^{\rm GR}\sqrt{\frac{L_{\rm TRGB}}{L_{\rm TRGB,\, GR}}}.
\end{equation}
These results indicate that comparing TRGB distances with other indicators, in particularly Cepheids which we have already demonstrated may be unscreened for some parameter choices, is a promising test of this mechanism. We perform this test quantitatively in a companion paper \cite{Desmond:2019ygn}.

\section{White Dwarfs and Type Ia Supernovae}
\label{sec:WD}

Type Ia supernovae are standardizable candles, which has made them a powerful tool for cosmology. The progenitors of Type Ia supernovae are white dwarf stars that exceed their Chandrasekhar mass and undergo a thermonuclear run-away. The primary variable determining the peak luminosity of the light curve is the mass of nickel-56 available for the explosion. A simple approximation is that the white dwarf is composed entirely of $^{56}$Ni so that the mass of nickel is equal to the Chandrasekhar mass $M_{\rm Ch}$. White dwarfs close to the Chandrasekhar limit are very accurately described by a gas of relativistic particles, which have equation of state $P=K\rho^\frac{4}{3}$ i.e. they it is polytropic with index $n=3$ (see Appendix \ref{sec:poly} where we provide a brief introduction to polytropic stars). 

The Chandrasekhar mass is derived in Eq. \eqref{eq:Mch} where one can see that $M_{\rm Ch}\propto \GN^{-\frac32}$. The equivalent expression for unscreened white dwarfs can be found by rescaling $\GN\rightarrow\GN(1+\Delta G/\GN)$. The ratio of the two is
\begin{equation}
    \frac{M_{\rm Ch}(\Delta G)}{M_{\rm Ch}}=\left(1+\frac{\Delta G}{\GN}\right)^{-\frac32}.
\end{equation}
The na\"{i}ve expectation is then that unscreened type Ia supernovae have lower peak luminosities than their screened d\"{o}pplegangers owing to their lower Chandrasekhar masses, which implies less $^{56}$Ni. This conclusion is indeed na\"{i}ve. Ref. \cite{Wright:2017rsu} has studied the effects of changing the strength of gravity on type Ia supernovae. Using a semi-analytic model, they found that after one accounts for the variation of the mass of $^{56}$Ni with the total mass, and the standardization procedure where the light-curve is stretched to match a standard template, the ultimate effect of strengthening gravity is to enhance the peak luminosity. In the context of our model, this implies that the distance to unscreened supernovae is underestimated if one does not account for this enhancement. Fitting fig. 7 (left) of Ref. \cite{Wright:2017rsu}, one finds a relation
\begin{equation}
    \frac{L_{\rm peak}(\Delta G)}{L_{\rm peak,\, GR}}=\left(1+\frac{\Delta G}{\GN}\right)^C
\end{equation}
with $C\simeq1.46$. This formula can be used in future observational probes of (for example) screened baryon--dark matter interactions involving type Ia supernovae, either using observations of individual events or by comparing different distance ladder measurements.

\section{Low Mass Stars}
\label{sec:dwarf_stars}

Dwarf stars have proven to be powerful probes of gravity theories where the gravitational constant varies inside of extended objects \cite{Sakstein:2015zoa,Sakstein:2015aac,Crisostomi:2019yfo}. Brown dwarf stars are inert objects that are supported by Coulomb scattering pressure. They are not heavy enough for their cores to reach the temperatures and densities necessary for hydrogen fusion. Red dwarf stars are heavier objects that can achieve the central conditions necessary to ignite hydrogen burning, albeit on a chain that results in the net production of $^3$He rather than the PP chains (which result in $^4$He). The transition mass separating these two objects is known as the \emph{minimum mass for hydrogen burning}, $M_{\rm Min}$, and the relative simplicity of dwarf stars allows for its analytic calculation with few degeneracies. This has made it a powerful probe of cosmologically relevant modified gravity theories \cite{Sakstein:2015zoa,Sakstein:2015aac}.

The screening mechanism we have derived in this work allows for the possibility that stars in the outskirts our own galaxy are unscreened since the dark matter density falls below that in the solar neighborhood in these regions. For this reason, in this section we will study how changing the gravitational constant effects the properties of dwarf stars.

\subsection{Red Dwarfs and the Minimum Mass for Hydrogen Burning}

The derivation of the minimum mass for hydrogen burning is standard in the literature, so we will not repeat it here. Instead, we will simply focus on the underlying physics, referring the reader to \cite{Sakstein:2015zoa,Sakstein:2015aac,Crisostomi:2019yfo} for the analytic details. Stable hydrogen burning is achieved when the luminosity due to nuclear fusion in the stellar core balances the luminosity losses from the photosphere. If the gravitational constant increases to values larger than $\GN$ the star will become more compact, raising the central temperature and density. One therefore expects the minimum mass for hydrogen burning to be smaller in unscreened stars than screened stars, or, said another way, it is possible that brown dwarfs in the solar neighborhood could be red dwarfs in the galactic outskirts. This could potentially change the shape of the Hertzsprung--Russell diagram measured in globular clusters in the galactic outskirts.

\begin{figure*}[t]
    \centering
    \includegraphics[width=0.45\textwidth]{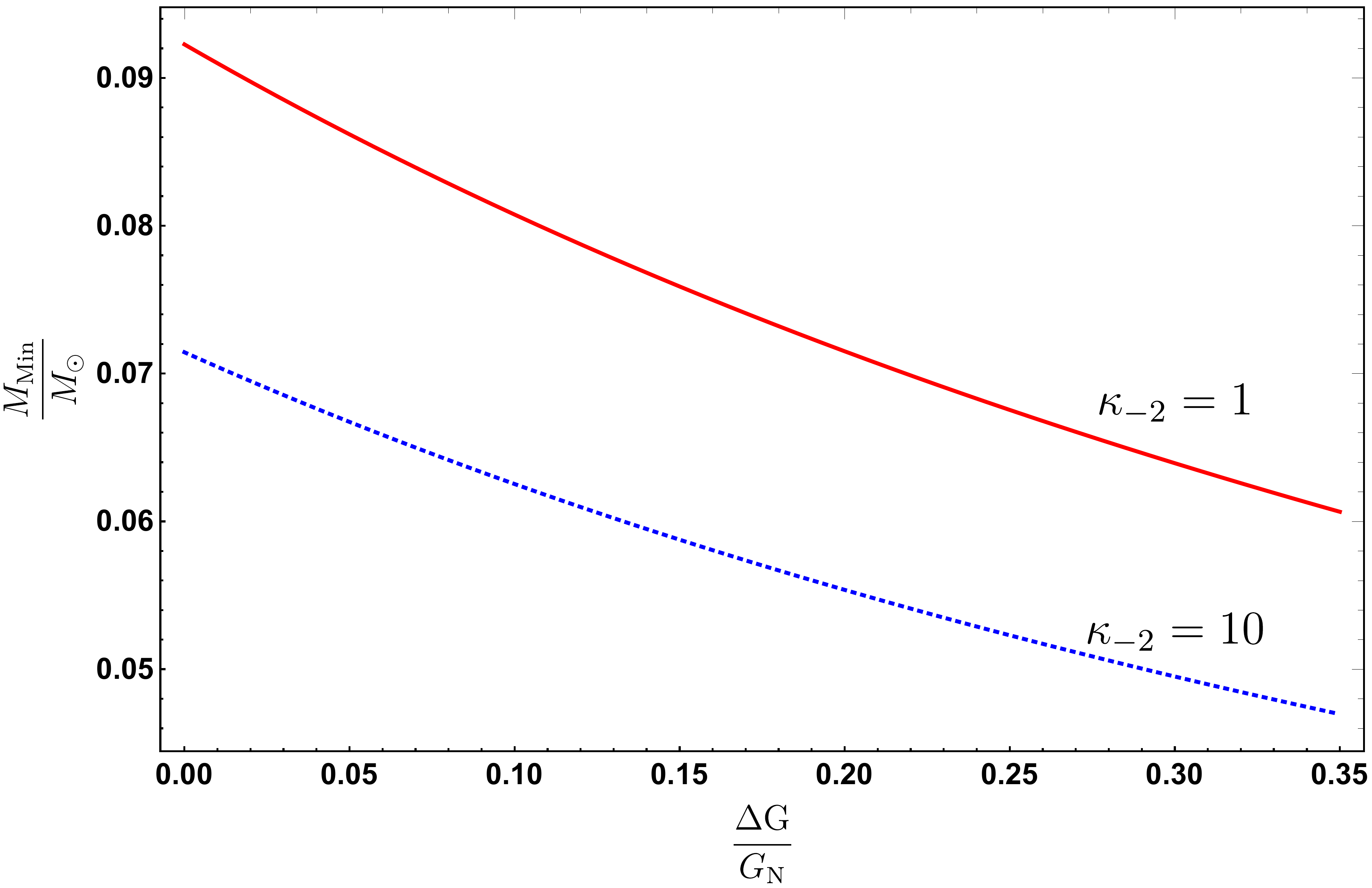}
    \includegraphics[width=0.45\textwidth]{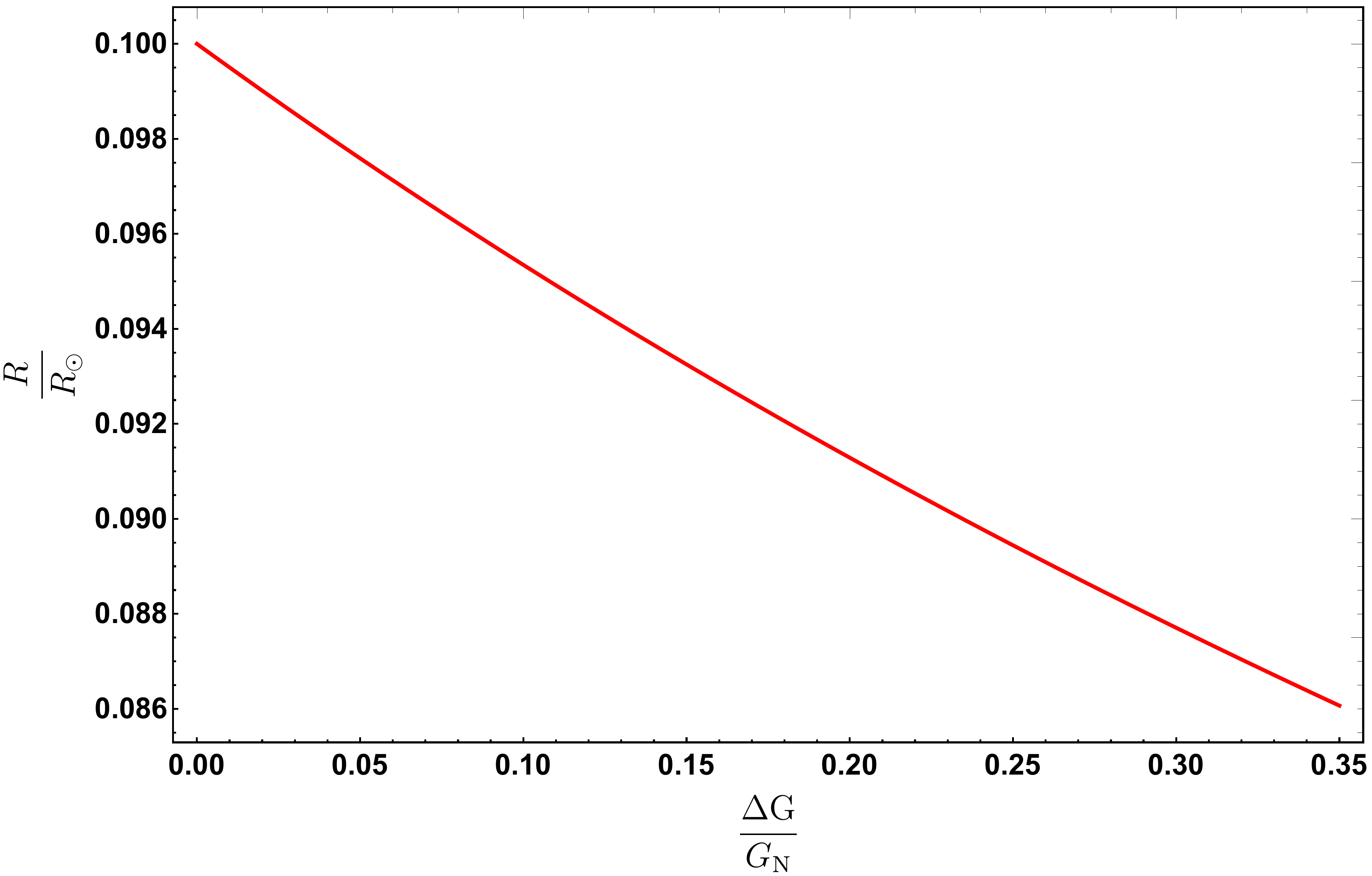}
    \caption{\emph{Left}: The minimum mass for hydrogen burning as a function of $\Delta G/\GN$ for two representative opacities given in the figure. \emph{Right}: The (mass-independent) radius of brown dwarf stars as a function of $\Delta G/\GN$.  }
    \label{fig:dwarf}
\end{figure*}

In order to quantify this effect, we will use the formula derived by Ref. \cite{Crisostomi:2019yfo}, Appendix B2:
\begin{equation}
    \frac{M_{\rm Min}}{M_\odot}= 0.092\kappa_{-2}^{-0.111}\left(1+\frac{\Delta G}{\GN}\right)^{-1.398},
\end{equation}
where $\kappa_{-2}$ is the Rosseland mean opacity measured in units of $10^{-2}$ g cm$^{-2}$. This is the only source of uncertainty in this equation and it is not very significant. We plot $M_{\rm min}$ as a function of $\Delta G/\GN$ in the left panel of Fig. \ref{fig:dwarf}. In GR, the minimum mass for a red dwarf is $M_{\rm Min}\sim 0.08 M_\odot$. From the figure one can see that, depending on the opacity and $\Delta G$, in unscreened regions of the galaxy stars with masses as low as $0.05 M_\odot$ could be hydrogen-burning red dwarfs.

\subsection{Brown Dwarfs and the Radius Plateau}

Stars with masses smaller than $M_{\rm min}$ are brown dwarfs supported by Coulomb scattering pressure. Objects supported by Coulomb scattering have a fixed radius that is independent of their mass \cite{Sakstein:2015aac}. This leads to a distinctive \emph{radius plateau} in the Hertzsprung--Russell diagram for low mass objects. In GR, the radius is given by $R\sim 0.1 R_\odot$ but the formula used to derive this contains a factor of $\GN^{-1/2}$ \cite{Sakstein:2015aac} so that the radius of unscreened stars scales as
\begin{equation}
    R=0.1\left(1+\frac{\Delta G}{\GN}\right)^{-\frac12} R_\odot.
\end{equation}
Physically, increasing Newton's constant makes the star more compact. We plot the radius of brown dwarf stars as a function of $\Delta G/\GN$ in the right panel of Fig. \ref{fig:dwarf}. The predicted decrease in radius plateau shown in this figure can potentially be tested with the brown dwarfs observed by GAIA provided one can measure the radius. If this is the case, one could plot the brown dwarf radius as a function of distance from the galactic center: in our model one might expect to see a negative correlation.

\section{Dynamical vs. Lensing Mass}
\label{sec:ML}

In this section we discuss potential tests of the dark matter density-dependence of the PPN parameter $\gamma(\rdm)$ predicted in Eq. \eqref{eq:gammarho}, focusing on strong and weak gravitational lensing. Both of these rely on the fact that the dynamical mass of an object $M_{\rm Dyn}$ (measured using dynamical or kinematic tracers) differs from the lensing mass $M_{\rm Lens}$ (measured using gravitational lensing observations) according to
\begin{equation}
    \label{MvsL}
    M_{\rm Lens}=\frac{1+\gamma(\rdm)}{2}M_{\rm Dyn}.
\end{equation}

\subsection{Strong Gravitational Lensing}

For this test, typically applicable to clusters and massive elliptical galaxies, the lens mass is determined from the radius of the Einstein ring or multiple image locations. (Note that the dark matter density at the Einstein radius and in galaxy clusters is higher than locally.)  
The dynamical mass is determined from the velocity dispersion of stars in elliptical galaxies and galaxy clusters, or from X-ray emission or the Sunyaev-Zel’dovich signal of the hot gas assumed to be in hydrostatic equilibrium. The most stringent extragalactic bound, $\gamma=0.97\pm0.09$ in the galaxy ESO 325-G004, was reported by Ref. \cite{Collett:2018gpf}. Similarly, reference \cite{Schwab:2009nz} report $\gamma=1.01\pm0.05$. 
This bound is unlikely to be useful for constraining this screening mechanism because the dark matter density at the location of the lens is likely larger than the screening threshold for the Solar System (the central velocity dispersion of ESO 325-G004, $371$ km/s, implies a dark matter density a few times that of the MW), but similar tests in less dense dark matter environments may be possible.

\subsection{Weak Gravitational Lensing}

Tests using weak lensing may be more fruitful since one can stack many galaxies to make measurements in less dense environments. Several variants of this test have already been performed in the context of testing gravity. On cluster scales, one can use the X-ray surface brightness as a probe of the dynamical mass, and $30\%$ bounds on $\gamma$ have been obtained \cite{Terukina:2013eqa,Wilcox:2015kna,Sakstein:2016ggl,Pizzuti:2016ouw}. Other possibilities on cosmological scales are to cross-correlate weak lensing measurements with redshift space distortions \cite{Simpson:2012ra} or peculiar velocity measurements \cite{Song:2010fg}, or to constrain the $E_G$ statistic \cite{Blake:2015vea,Singh:2018flu}. Applying these to our model would require deriving the equations for linear cosmological perturbations, which will depend on the fundamental theory functions, as well as more precise characterisation of the relevant dark matter densities.

\section{Summary and Conclusions}
\label{sec:concs}

In this section we summarize our results, discuss directions for future work, and conclude.

\subsection{Summary of the Paper}

In this paper we have demonstrated that baryon--dark matter interactions can give rise to a new screening mechanism. While Ref. \cite{Berezhiani:2016dne} speculated that such a mechanism exists, we have derived it here in detail. We found that there are two consequences of the screening mechanism: the value of both Newton's constant and the PPN parameter $\gamma$ (which describes the relative motion of light and matter) become functions of the local dark matter density. The screening mechanism is present whenever the fundamental functions appearing in the theory are such that the baryon--dark matter interactions vanish in high dark matter densities but become important at lower density. This is what one would expect from any model where dark energy emerges at late times (low cosmological dark matter densities) from such interactions. 

Lacking canonical forms for these functions, we adopted a simple approach to understanding the screening mechanism and exploring its observational consequences. The Milky Way must be screened for the theory to pass local tests of GR so we expect that the underlying parameters of any viable theory are such that objects are only unscreened ($G(\rdm)>\GN$) in densities smaller than that at the position of the Solar System. We also assumed that at densities below this threshold, $G(\rdm)$ is larger than in the Solar System by a constant amount, i.e. $G(\rdm)=(1+\Delta G/\GN)\GN$. This allowed us to understand the observational consequences using a one-parameter description of unscreened objects. A more complex dependence on $\rdm$ could lead to $\Delta G/\GN$ for astrophysical objects varying between different halos or even within a given halo. It is unlikely that this parameter will vary over the scale of stellar objects, however, since in conventional dark matter models these do not induce large variations in $\rdm$. The results of this work can then be considered general provided that one bears in mind that $\Delta G/G$ may need to be varied on an object-by-object basis even within a single halo. 

We next discussed several observational tests. Under our simple model this is tantamount to understanding how the properties of astrophysical object are altered when one increases Newton's constant by fixed amount\footnote{A similar phenomenon occurs in degenerate higher-order scalar tensor theories (DHOST) \cite{Hirano:2019scf,Crisostomi:2019yfo}. In this case the value of Newton's constant is altered inside astrophysical bodies due to a breaking of the Vainshtein mechanism (there is no dependence on the dark matter density). The phenomena discussed in this work could therefore also be used to constrain DHOST theories.}. Since the Solar System must be screened, the most likely tests would be extragalactic. For this reason, we focused on objects in other galaxies whose properties can be measured --- Cepheid variable stars and type Ia supernovae. 
The Cepheid period--luminosity relation in particular is a powerful tool for constraining screening mechanisms \cite{Jain:2012tn,Sakstein:2013pda}. By running a suite of numerical simulations using a version of the stellar structure code MESA \cite{2011ApJS..192....3P,2015ApJS..220...15P,Paxton:2017eie,2019arXiv190301426P} that was modified to change the value of Newton's constant, we derived the modification to the PLR as a function of $\Delta G/\GN$: under a fifth force we find a decrease in pulsation period and overall increase in luminosity. We provide the coefficients describing the (mass-dependent) luminosity effect in Table \ref{tab:Bs}. We found that the distance to unscreened galaxies would be underestimated if one assumed the GR PLR. We also discussed how type Ia supernovae, which are standardizable candles, would behave if they are unscreened, finding that the light curve peak luminosity (after stretch-correction) is enhanced.

Using analytic and numerical techniques, we demonstrated that unscreened main-sequence stars are, as expected, more luminous at fixed mass and metallicity than Newtonian or screened stars, although degeneracies with the metallicity will likely make tests using individual objects difficult. The one exception to this is stars observed in our own galaxy. In particular, globular clusters in the outskirts of the Milky Way may be unscreened due to the smaller dark matter density. There are fewer degeneracies associated with these objects since the metallicity and age can be determined. The luminosity at the tip of the red giant branch was found to be smaller than the GR value, implying that the application of the GR formula will overestimate the true distance. It is thus possible to test our predictions by comparing the TRGB and Cepheid distances to unscreened galaxies, as we do in a companion paper \cite{Desmond:2019ygn}.

As ancillary results, we studied how low mass red and brown dwarf stars respond to an increased Newton's constant. These objects cannot be observed in other galaxies but it is possible that Milky Way dwarfs are unscreened if they are sufficiently far from the solar neighborhood that the dark matter density is sufficiently low. We found that the minimum mass for hydrogen burning, which separates red from brown dwarfs, decreases for $G>G_\text{N}$, so that objects with masses as low as $0.05M_\odot$ could be red dwarfs, compared with GR where the threshold is $0.08 M_\odot$. We also found that the radius plateau in the brown dwarf Hertzsprung--Russell diagram would shift to smaller radii. Given that GAIA is able to observe these objects, it is possible for these predictions to be tested in the near future. In more complicated models than we have studied here it is possible that $G(\rdm)<\GN$ in some galaxies. One can use our results to calculate the observable consequences of this, e.g. main-sequence stars would be less luminous than in GR and the minimum mass for hydrogen burning would be larger.

Finally, we discussed tests of the predicted dark matter density-dependence of the PPN parameter $\gamma$. Measuring this in other galaxies requires determining both dynamical and lensing mass. Such mass vs. light tests have become a standard probe of modified gravity theories, and both weak and strong lensing systems could provide interesting bounds.

\subsection{Future Directions}
\label{sec:future}
We have performed a preliminary study of the baryon--dark matter screening mechanism. Here we discuss future theoretical and observational directions. {Since the threshold for screening corresponds to the dark matter density at redshift $z\le 40$ we expect that at times earlier than this, the Universe evolves in an identical manner to the predictions of GR. For this reason, we only discuss late-time phenomenology. }

\begin{itemize}
    \item \textbf{Construction of a Canonical Model:} We have (deliberately) not chosen any specific form for the free functions describing the baryon--dark matter interactions in this work. Instead, we have treated $G$ and $\gamma$ as generic functions of $\rdm$. Choosing forms for these would allow for further tests using observables that are not described simply by a constant shift in $G$. It would be convenient if a canonical model could account for dark energy, but not essential because the screening mechanism may have interesting and potentially useful consequences on smaller scales.
    \item \textbf{Gravitational Waves:} In this theory, gravitational waves move on geodesics of the Einstein frame metric $g_\nm$ but light moves on geodesics of the Jordan frame metric $\tg_\nm$, which implies that they may move at different speeds in general. The simultaneous detection of light and gravitational waves from a neutron star merger event (GW170817  \cite{TheLIGOScientific:2017qsa}) has constrained any difference to be at the $10^{-15}$ level, which has placed strong bounds on modified gravity theories   \cite{Sakstein:2017xjx,Ezquiaga:2017ekz,Creminelli:2017sry,Baker:2017hug}. This deviation could be used to constrain the free functions of the present theory. It is highly likely that the combination of parameters that sets the speed of gravitational waves is different from the combinations setting $G(\rdm)$ and $\gamma(\rdm)$, so this strong bound is not necessarily debilitating. It is also possible to construct theories from first principles where the speed of light and gravity are identical \cite{Crisostomi:2017pjs,Copeland:2018yuh,Kase:2018aps,Kobayashi:2019hrl}. One should check the manner in which such bounds apply, and also that the theory is not being applied outside the range of validity of effective field theory \cite{deRham:2018red}.
    \item \textbf{Distance Indicator Comparisons:} Comparing screened and unscreened distance estimates to the same galaxy has yielded strong bounds on chameleon (and similar) theories \cite{Jain:2012tn}. In this work we have studied both Cepheids and TRGB distance indicators, finding that both are unscreened with the former under-estimating the true distance and the latter over-estimating it. A test of this mechanism along the lines of \cite{Jain:2012tn} could therefore yield important constraints on $\Delta G/\GN$. We perform this test in a companion paper \cite{Desmond:2019ygn}.
    \item \textbf{Binary Pulsars:} Theories that include disformal couplings may predict different values of Newton's constant for non-relativistic objects relative to gravitational waves. In particular, the orbital decay of binary pulsars is sensitive to $G_{\rm GW}$ and $c_T/c$, the speed of gravitational waves relative to the speed of light discussed above \cite{Jimenez:2015bwa}. Constraints are at the $10^{-2}$ level, so a calculation of $G_{\rm GW}$ once a canonical model is specified would be highly constraining. 
    \item \textbf{The Hubble tension:} The discrepancy between the Planck central value of the Hubble constant and the value inferred using the distance ladder is currently $4.4\sigma$ \cite{Riess:2019cxk}, which is strong evidence for new physics affecting one or both of these probes. Unscreening galaxies that calibrate the supernova magnitude--redshift relation could provide a resolution. This is the focus of a companion paper \cite{Desmond:2019ygn}.
    \item \textbf{Baryon fraction in collapsed objects:} The baryon fraction in galaxies and clusters is time-dependent if dark matter and baryons redshift at different rates \cite{Goncalves:2015eaa}. In this theory, baryons redshift as $a^{-3}$ but dark matter will redshift at a different rate (in a manner dependent on the choice of fundamental theory functions) since it moves on geodesics of a different (Einstein frame) metric. In other words, dark matter and baryons inhabit effective space-times that are expanding at different rates. The theory may therefore be tested by measuring the baryon fraction as a function of redshift. 
    \item \textbf{Cosmological Tests:} {Once a canonical model is specified, one could derive the equations of motion for the background cosmology and linear cosmological perturbations and use these to constrain the theory using current data sets, for example, CMB, BAO, and LSS data. The non-standard redshifting of dark matter at late times (discussed in the previous bullet point) implies that the amount of dark matter required in the early Universe may differ from the $\Lambda$CDM value. Thus, further tests using the relic abundance of dark matter and BBN may be possible.  }
    \item \textbf{The Integrated Sachs-Wolfe Effect:} The theory predicts that photons interact with dark matter in a manner that causes them to feel the fifth force. This implies that the integrated Sachs-Wolfe (ISW) effect will differ from the GR prediction. Other theories that make similar predictions have been tightly constrained by ISW observations \cite{Barreira:2012kk,Renk:2017rzu}.
    \item \textbf{Tests of the weak equivalence principle between baryons and dark matter:} Unscreened baryons feel a fifth force whereas dark matter does not so that the weak equivalence principle (WEP) is violated. Extragalactic tests of the WEP could constrain this screening mechanism. Indeed, such tests have been fruitful in the context of chameleon screening \cite{Jain:2011ji,Vikram:2013uba,Vikram:2014uza,Desmond:2018sdy,Desmond:2018euk,Desmond:2018kdn}, although in that case it is dark matter that is typically unscreened so one would need to derive the predictions for our theory separately.
    \item \textbf{Two-body calculations:} The analysis we have performed in this work considered a single isolated object. It would be interesting to extend this to two-body systems. In some cases, a violation of the WEP can emerge \cite{Hiramatsu:2012xj,Sakstein:2017pqi}. The baryon--dark matter interaction is described by a disformal transformation  \cite{Sakstein:2014aca,Sakstein:2014isa,Sakstein:2015jca}, which are known to exhibit novel effects in many-body systems \cite{Brax:2018bow,Kuntz:2019zef,Brax:2019tcy}.
    \item \textbf{Tests in the strong-field regime:} Strong-field tests of gravity are a powerful probe of alternatives to GR. Once a canonical model is specified, it would be straightforward to calculate the equivalent of the Tolman-Oppenheimer-Volkov equations for the theory and hence derive the properties of neutron stars in unscreened galaxies. 
    \item \textbf{Dark matter detection experiments:} {The authors of reference \cite{Berezhiani:2016dne} noted that, similarly to how Newton's constant is dark matter density-dependent, Fermi's constant is likewise. This may have implications for dark matter experiments (both direct and indirect), although it is straightforward to see that modifications are screened in high density environments so the bounds are weak. We  refer the reader to \cite{Berezhiani:2016dne} for the details.  }
    \end{itemize}

\subsection{Conclusion}

The new screening mechanism that we have studied in this work derives from a dark energy model that is not yet constrained, or even well characterized. The scarcity of dark energy models that are not ruled out or fine-tuned makes this model an interesting candidate for future study. The observational consequences we have derived here could form the basis for new astrophysical probes of the mechanism, which should be supplemented with cosmological tests, tests in the strong-field regime, and other tests described above. The construction of a canonical model with which to benchmark constraints would be a useful first step.

We have suggested several future directions, both theoretical and observational, that could lead to a deeper understanding and further bounds on this screening mechanism. These may ultimately prove too constraining to allow the theory to account for dark energy, as is the case with other screening mechanisms. Even if this were to happen the screening mechanism would remain of interest in its own right, and regions of parameter space where $G$ varies between galaxies but the theory does not behave as dark energy may still provide interesting intermediate-scale phenomenology.

Finally, we note that we have focused on devising novel probes of the theory's prediction that Newton's constant varies as a function of the local dark matter density. There are other theories (either fundamental or phenomenological) where $G$ is spatially-varying in a manner that is correlated with physical quantities. The tests we have devised in this work apply equally to these theories, provided one correlates $G$ correctly with the variables that govern its variation.

\section*{Acknowledgements}

We are grateful for discussions with Koushik Sen, Radek Smolec, and the wider MESA community for answering our various MESA related questions. We would like to thank Justin Khoury for several enlightening discussions. JS is supported by funds made available to the Center for Particle Cosmology by the University of Pennsylvania. HD is supported by St John's College, Oxford, and acknowledges financial support from ERC Grant No 693024 and the Beecroft Trust. BJ is supported in part by the US Department of Energy grant DE-SC0007901 and by NASA ATP grant NNH17ZDA001N.

\bibliography{ref}

\appendix

\section{Stellar Structure Models}
\label{sec:poly}

In this section we briefly review polytropic equations of state and spherically symmetric polytropic stars in GR, which are now standard in the literature relating screening mechanisms to stellar structure. Further details can be found in Refs. \cite{Davis:2011qf,Koyama:2015oma,Sakstein:2015oqa}.\footnote{See also \href{http://www.jeremysakstein.com/astro_grav_2.pdf}{http://www.jeremysakstein.com/astro\_grav\_2.pdf}.}

\subsection{Polytropic Stars}

A polytropic equation of state is one where the pressure
\begin{equation}\label{eq:polyeos}
    P=K\rho^\gamma=K\rho^{\frac{n+1}{n}},
\end{equation}
where $K$ is a constant and $n$ is referred to as the \emph{polytropic index}. When used in conjunction with the hydrostatic equilibrium equation
\begin{equation}\label{eq:HSE}
    \frac{\dd P}{\dd r}=-\frac{\GN M(r)\rho(r)}{r^2}
\end{equation}
and the equation of mass conservation
\begin{equation}\label{eq:dmdr}
    \frac{\dd M(r)}{\dd r^2}=4\pi r^2\rho(r),
\end{equation}
the equations of stellar structure become self-similar so that all dimensionful quantities can be scaled out, and the equations can be written purely in terms of dimensionless variables. This is achieved as follows. First, define the dimensionless quantities
\begin{equation}\label{eq:LEV}
   \xi=\frac{r}{r\ccc},\quad \rho=\rho\ccc\theta(\xi)^n,\quad r\ccc^2=\frac{(n+1)P\ccc}{4\pi G \rho\ccc^2},
\end{equation}
where $P\ccc$ and $\rho\ccc$ are the central pressure and density respectively. Using Eq. \eqref{eq:polyeos}, one has $P=\rho\ccc\theta(\xi)^{n+1}$ so that the function $\theta(\xi)$ completely characterizes the pressure and density profiles of the star in terms of the dimensionless radial coordinate $\xi$. (In fact, $\theta(\xi)$ is the dimensionless temperature because $T=T\ccc\theta(\xi)$.) Combining the relations \eqref{eq:LEV} with the stellar structure Eqs. \eqref{eq:HSE} and \eqref{eq:dmdr} one finds the Lane-Emden equation
\begin{equation}\label{eq:LaneEmden}
    \frac{1}{\xi^2}\frac{\dd}{\dd\xi}\left(\xi^2\frac{\dd\theta(\xi)}{\dd\xi}\right)=-\theta(\xi)^n.
\end{equation}
This can be solved for a given value of $n$ either numerically, or, in the cases $n=0,\,1,\,5$, analytically. The boundary conditions are $\theta(0)=1$ ($\rho(r=0)=\rho\ccc$) and $\theta'(0)=0$ (by virtue of spherical symmetry $\dd P/\dd r|_{r=0}=0$, which can be seen from Eq. \eqref{eq:HSE} by setting $M(0)=0$). The radius of the star is the point where $P(R)=0$. One can find this by integrating Eq. \eqref{eq:LaneEmden} to the point $\xi_R$ such that $\theta(\xi_R)=0$. The radius is then
\begin{equation}\label{eq:radLE}
R=r\ccc\xi_R. 
\end{equation}
Substituting Eq. \eqref{eq:LaneEmden} into Eq. \eqref{eq:dmdr} and integrating from $\xi=0$ to $\xi=\xi_R$ one finds an expression for the stellar mass
\begin{equation}
    \label{eq:massE}
    M=4\pi r\ccc^3\omega_R\rho\ccc=4\pi\omega_R\left[\frac{(n+1)K}{4\pi \GN}\right]^{\frac32}\rho\ccc^{\frac{3-n}{2n}},
\end{equation}
where
\begin{equation}
    \omega_R\equiv -\xi_R^2\left.\frac{\dd\theta}{\dd\xi}\right\vert_{\xi_R}
\end{equation}
and we have used Eq. \eqref{eq:LEV}. The analytic models studied in this work can all be described by $n=3$ polytropic models. In this case, according to Eq. \eqref{eq:massE} the mass is independent of the central density. For systems comprised of gasses that are nearly relativistic ($\gamma=4/3,\,n=3$), this represents the limiting, Chandrasekhar mass:
\begin{equation}\label{eq:Mch}
    M_{\rm Ch}=4\pi r\ccc^3\omega_R\rho\ccc=4\pi\omega_R\left[\frac{K}{\pi \GN}\right]^{\frac32}.
\end{equation}

\vspace{10mm}

\subsection{The Eddington Standard Model}

Here we briefly derive the Eddington standard model which is used in Sec. \ref{sec:ESM}. This model makes several simplifying assumptions in order to model main-sequence stars analytically. The pressure support is assumed to be due to two different processes: the motion of the gas, assumed to be ideal
\begin{equation}\label{eq:Pgas}
    P_{\rm gas}=\frac{k_{\rm B} \rho T}{\mu m_{\rm H}},
\end{equation}
where $\mu$ is the mean molecular mass, and radiation pressure
\begin{equation}\label{eq:Prad}
    P_{\rm rad}=\frac{1}{3}a T^4.
\end{equation}
The relative importance of each contribution is quantified by 
\begin{equation}
    \label{eq:betadef}
    \beta\equiv\frac{P_{\rm gas}}{P}.
\end{equation}
Using Eqs. \eqref{eq:Pgas} and \eqref{eq:Prad}, one finds
\begin{equation}\label{eq:s}
    \frac{T^3}{\rho}=3a\frac{k_{\rm B}}{\mu m_{\rm H}}\frac{1-\beta}{\beta}.
\end{equation}
The quantity $T^3/\rho$ is the specific entropy density. The key assumption of the Eddington standard model is that this is constant throughout the star. Equation \eqref{eq:s} then implies that $\beta$ is constant. With this assumption, the total pressure is 
\begin{equation}
    \label{eq:Ptot}
    P=P_{\rm rad}+P_{\rm gas}=K(\beta)\rho^{\frac{4}{3}},
\end{equation}
with
\begin{equation}
    \label{eq:Kbeta}
    K(\beta)=\left(\frac{3}{a}\right)^{\frac13}\left(\frac{k_{\rm B}}{\mu m_{\rm H}}\right)^{\frac43}\left(\frac{1-\beta}{\beta^4}\right)^{\frac13}.
\end{equation}
We can eliminate $K$ in favor of the mass using Eq. \eqref{eq:massE}, to find
\begin{equation}\label{eq:Eddquartic}
    \frac{1-\beta}{\beta^4}=\left(\frac{M}{M_{\rm Edd}}\right)^2,
\end{equation}
where
\begin{equation}\label{eq:MEDD}
    M_{\rm Edd} = \frac{4\omega_R}{\sqrt{\pi} \GN^{\frac32}}\left(\frac{3}{a}\right)^{\frac12}\left(\frac{k_{\rm B}}{\mu m_{\rm H}}\right)^{2}
\end{equation}
is the \emph{Eddington Mass}. Equation \eqref{eq:Eddquartic} is known as \emph{Eddington's quartic equation}. It's solution gives $\beta (M)$. The key quantity we are interested in is the stellar luminosity, which can be found by inserting $P=(1-\beta(M))P_{\rm rad}$ into the hydrostatic equilibrium equation \eqref{eq:HSE} and substituting this into the equation of radiative transfer:
\begin{equation}
    \frac{\dd T}{\dd r} = -\frac{3}{4a}\frac{\kappa(\rho,T)}{T^3}\frac{\rho L}{4\pi r^2},
\end{equation}
where $\kappa (\rho,T)$ is the opacity. Making the further assumption that $\kappa$ is constant, one finds
\begin{equation}\label{eq:LEDD}
    L(M)=\frac{4\pi(1-\beta(M))\GN M}{\kappa}.
\end{equation}
This relates the luminosity to the mass and $\GN$, and is the equation we use in Sec.~\ref{sec:ESM}.

\end{document}